\documentclass[twocolumn]{aa}
\usepackage{graphicx}
\usepackage{txfonts}

\begin{document}

\title{Luminosity and mass functions of the three main sequences of
  the globular cluster NGC 2808\thanks{Based on observations with the
  NASA/ESA {\it Hubble Space Telescope}, obtained at the Space
  Telescope Science Institute, which is operated by AURA, Inc., under
  NASA contract NAS 5-26555, under the programs GO-9899 and GO-10922}}

\author{
A.\,P.\, Milone\inst{1,2,3},
G.\, Piotto\inst{3,4},
L.\, R.\, Bedin\inst{4,5},
S.\, Cassisi\inst{6},
J.\, Anderson\inst{5},
A.\, F.\, Marino\inst{7},
A.\, Pietrinferni\inst{6}, 
A.\, Aparicio\inst{1,2}
}

\offprints{A.\ P.\ Milone}

\institute{
           Instituto de Astrof\`\i sica de Canarias, E-38200 La
           Laguna, Tenerife, Canary Islands, Spain\\
           \email{milone@iac.es, aparicio@iac.es}   
           \and
           Department of Astrophysics, University of La Laguna, E38200
           La Laguna, Tenerife, Canary Islands, Spain
           \and
           Dipartimento  di   Astronomia,  Universit\`a  di Padova,
           Vicolo dell'Osservatorio 3, Padova, I-35122, Italy\\
           \email{giampaolo.piotto@unipd.it}          
           \and
           INAF-Osservatorio Astronomico di Padova, Vicolo dell'Osservatorio 5,
           Padova I-35122, Italy
           \and
           Space Telescope Science Institute, 3700 San Martin Drive,
           Baltimore, MD 21218, USA\\
           \email{bedin@stsci.edu, jander@stsci.edu}
           \and
           INAF-Osservatorio Astronomico di Collurania, via Mentore
           Maggini, 64100 Teramo, Italy\\
           \email{cassisi@oa-teramo.inaf.it,pietrinferni@oa-teramo.inaf.it}
           \and
           Max Plank Institute for Astrophysics, Postfach 1317, 85741, Garching,
           Germany\\
           \email{amarino@MPA-Garching.MPG.DE}
           }

\date{Received Xxxxx xx, xxxx; accepted Xxxx xx, xxxx}
%
\abstract{ High-precision {\it HST} photometry has revealed
that the globular cluster (GC) NGC 2808 hosts a triple main sequence
(MS) corresponding to three stellar populations with different helium
abundances. We carried out photometry on ACS/WFC {\it HST} images
of NGC 2808 with the main purpose of measuring the luminosity function
(LF) of stars in the three different MSs, and the binary fraction in the
cluster.  We used isochrones to transform the observed LFs into mass functions
(MFs).

 In NGC 2808  we estimate that the fraction of binary systems is $f_{\rm bin}
\simeq 0.05$, and find that the three MSs have very similar LFs.  The
slopes of the corresponding MFs are $\alpha=-1.2\pm0.3$ for the red
MS, $\alpha = -0.9 \pm 0.3$ for the middle MS, and $\alpha = -0.9 \pm
0.4$ for the blue one, the same, to within the errors.
 There is marginal evidence of a MF flattening for masses
${\mathcal M}\le 0.6 {\mathcal M}_{\rm \odot}$ for the the reddest
(primordial) MS.
These results represent the first direct measurement of the
present-day MF and LF in distinct stellar populations of a GC, and
provide constraints on models of the formation and
evolution of multiple generations of stars in these objects.

}
\titlerunning{Luminosity and mass functions of the three main sequences of NGC 2808}
\authorrunning{Milone et al.}

\maketitle

\section{Introduction}
\label{introduction}
The recent discoveries of multiple populations in globular clusters
(GCs) have demonstrated that these stellar systems are not as simple
as we first imagined. The findings of multiple evolutionary
sequences in the colour magnitude diagram (CMD) of several GCs has
been interpreted as sound evidence of
sub-populations characterised by different ages and/or chemical
abundances (Piotto et al.\ 2005, 2007; Milone et al.\ 2008; Cassisi et
al.\ 2008; Marino et al.\ 2009 and references therein), and 
has demonstrated that these systems are not single stellar populations.
In this context, helium has been proposed as a key element to
interpret both the horizontal-branch (HB) morphology and the multiple
and spread main sequence (MS) observed in some GCs.

The strongest indication of He-rich stellar
populations in GCs is that found for NGC 2808.  While spectroscopy of its
red giants has long shown it to have nearly uniform iron content
(Carretta et al.\ 2006), its main sequence is divided into three distinct 
sequences (D'Antona et al.\ 2005, Piotto et al.\ 2007), which have been 
interpreted as successive episodes of star formation.

Apart from its triple MS, NGC 2808 also displays additional observational
evidence of multiple stellar populations.
Its HB is greatly extended blueward, and is well populated to both the
blue and the red of the instability strip (see Dalessandro et al.\
2011 and references therein). The distribution of stars along the HB
is multimodal (Sosin et al.\ 1997, Bedin et al.\ 2000) with three
significant gaps, one of these gaps being at the colour of the RR
Lyrae instability strip. Even though the HB is well populated
to both the blue and the red of the instability strip, very few RR
Lyrae stars have been identified in NGC 2808. The other two gaps accur
along the blue extension of the HB, and delimit three distinct segments.

The red-giant branch (RGB) also exhibits evidence of non-singular
populations. The CMDs shown by 
Yong et al.\  (2008) 
and Lee et al.\ (2009) revealed a large colour spread among RGB stars that cannot be
attributed to photometric errors.  Furthermore, an analysis of
medium-high-resolution spectra of 122 RGB stars have revealed an
extended Na-O anti-correlation in NGC 2808 (Carretta et al.\ 2006)
with the presence of three distinct groups of O-normal (peak at [\rm
O/Fe]$=+0.28$, including $61\pm 7\%$ of the stars), O-poor (peak at
[\rm O/Fe]$=-0.21$, including $22\pm 4\%$ of the stars), and
super-O-poor (peak at [\rm O/Fe]$=-0.73$, including $17\pm 4\%$ of the
stars) stars.  Bragaglia et al.\ (2010a) 
detected evidence of sub-populations with
distinct initial He contents along the RGB of NGC 2808 by using
various evolutionary features such as the dependence of the mean RGB
$T_{\rm eff}$ on the He abundance.

On the basis of their relative number counts, Piotto et al.\ (2007)
attempted to associate the three MSs with both the three HB segments
defined by Bedin et al.\ (2000), and the three different
O-content groups detected by Carretta et al.\ (2006).

As almost all cluster stars have the same iron content (Carretta et
al.\ 2006), a multimodal distribution of He abundances seems to be the
only way to account for both the complex HB and the multiple MS
(D'Antona et al.\ 2005, Dalessandro et al.\ 2011), and is consistent
with the known abundance pattern of heavy elements. Following this
scenario, the population associated with the red MS (rMS) has a nearly
primordial helium, while stars in the middle (mMS) and blue MS (bMS)
are formed from the ejecta produced by an earlier stellar generation
in the complete CNO and MgAl cycle and are He-enhanced ($Y\sim0.32$
and $Y\sim0.38$, respectively).

The suitability of this scenario was confirmed by Bragaglia et
al.\ (2010b) who measured the chemical abundances of one star on the rMS
and one on the bMS and found that the latter shows an enhancement of
N, Na, and Al and a depletion of C and Mg, as one would expect from
material polluted by a first generation of massive stars.

The idea that the multiple stellar populations discovered in many GCs
can be related to the HB morphology is strongly supported by the
spectroscopic study of Marino et al.\ (2011) of the GC NGC 6121
(M4). These authors measured oxygen and sodium directly in HB stars
and found that the blue HB consists of O-poor (Na-rich, likely
He-rich) stars, in contrast to O-rich (Na-poor, likely He-normal) red
HB stars.

A multiple MS is also present in $\omega\,$Centauri where, a few
magnitudes below the MS turn off, the MS is divides into at least three
sequences (Anderson 1997, Bedin et al.\ 2004), with the bluer branch
hosting $\sim 30\%$ of the population (Bellini et al.\ 2009).  A
spectroscopic analysis based on data acquired using the FLAMES
spectrograph at the VLT by Piotto et al.\
(2005) revealed that, in contrast to expectations from evolution
models of stars with canonical abundances, the blue MS has twice the
metal abundance of the dominant red branch. In this case, we also note
that the only isochrones that are able to fit these combination of colors and
metallicity are greatly enriched in He (up to $Y\sim0.38$) with
respect to the old dominant component.

A divided or broad MS is not unique only to $\omega\,$Centauri
and NGC 2808, but is also present in at least two other GCs: 47
Tucanae (Anderson et al.\ 2009) and NGC 6752 (Milone et al.\ 2010).
In these cases, the colour spread has been tentatively attributed to
small variations in He ($\Delta Y \sim$ 0.02-0.03).

Considerable effort has been dedicated to interpreting these
observations.  It has been suggested that the bluer MSs are formed by
a distinct population of stars born from the He-enriched material
polluted by the ejecta of a first stellar generation. The most
appealing candidates are either intermediate-age asymptotic giant
branch (AGB) stars (e.g.\ Ventura et al.\ 2002, D'Antona et al.\ 2005)
or fast rotating massive stars (Decressin et al.\ 2007) although both
theoretical considerations (Renzini 2008) and observational evidence
(see the analysis on the Li-O correlation on the turn-off stars in
the GC NGC 6752 by Shen et al.\ 2010) seriously challenge the
rotating massive star scenario.  

In this work, we investigate, for the first time, the present-day
luminosity functions (LFs) and mass functions (MFs) of the three
stellar populations in NGC 2808.  On the basis of this study,
we also derive the fraction of binaries in these three sequences
and measure the fraction of stars in each population.

The paper is organised as follows.  In Sect.~\ref{sec:data}, we
describe the data and the data reduction.  The artificial star 
experiments and the procedure used to obtain the completeness are
illustrated in Sect.~\ref{AS}. In Sect.~\ref{sec:fiducial}, we derive
the fiducial of the three MSs and their colour spread. The LFs and the
fraction of binaries are calculated and discussed in Sect.~\ref{LUFU},
while Sect.~\ref{sec:MFs} is dedicated to the MFs of the three MSs.
Finally, a summary of the main results of this project and the
conclusions are given in Sect.~\ref{sec:conc}.

\section{Data and measurements}
\label{sec:data}
For this work, we used Hubble Space Telescope (\textit{HST\,}) data.
The images were collected with the Wide Field Channel on the Advanced
Camera for Surveys (ACS/WFC) under programs GO-9899 and GO-10922 and
are the same as those used in Piotto et al.\ (2007). This dataset consists of
ten exposures of $\sim 350$ s in F475W and six of the same exposure
time in F814W.  In addition, we also used a short-exposure image of 20
s in F475W and one of 10 s in F814W.

The photometric reduction was carried out using the software presented
and described in detail in Anderson et al.\ (2008). It consists of a
package that analyses all the exposures of each cluster simultaneously
to generate a catalogue of stars in the field of view. Stars are
measured independently in each image by using the point-spread function 
(PSF) 
library models from Anderson \& King (2006) with a spatially constant
perturbation tailored to each exposure that accounts for small focus
changes due to spacecraft breathing.

This routine allowed us to detect almost every star that could be
detected by eye and takes advantage of the many independent dithered
pointings of each scene and the knowledge of the PSF to avoid
including artifacts in the list. The ACS/WFC photometry was
calibrated into the Vega-mag system by following recipes in Bedin et
al.\ (2005) and using the zero points given in Sirianni et al.\
(2005).

\begin{figure*}[ht!]
   \centering
   \includegraphics[width=14 cm]{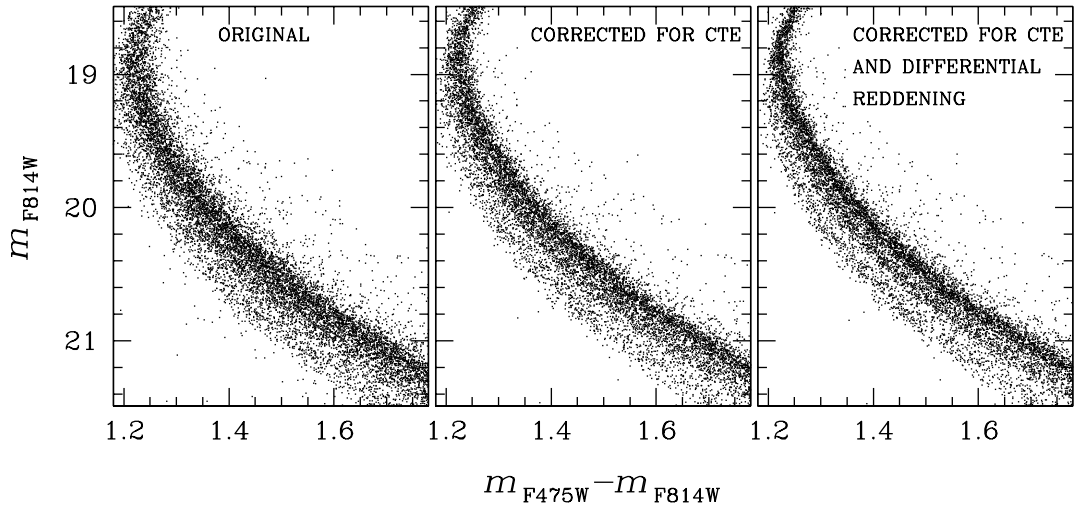}
   \caption{CMDs of NGC 2808 without any correction (\textit{left
       panel}), corrected for CTE (\textit{middle panel}), and
       corrected for both CTE and differential reddening
       (\textit{right panel}).  }
         \label{COMRED}
   \end{figure*}

We corrected the poor charge-transfer efficiency (CTE) in the ACS/WFC
data by using the procedure and the software developed by Anderson \&
Bedin (2010).  As an example, in the left and the middle panels of
Fig.~\ref{COMRED} we compare the ${\it m}_{\rm F814W}$ versus ${\it
m}_{\rm F475W}-{\it m}_{\rm F814W}$ CMD obtained from the original images
 with those from the images corrected for CTE.
%

\subsection{Differential reddening}
A visual inspection of the middle panel CMD of Fig.~\ref{COMRED}
reveals that NGC 2808 has broad MSs and sub-giant branch (SGB).  Some
or all of the photometric spread in color could in principle be due to
either differential reddening across the observed field or to
spatially dependent variations in the PSF encircled energy that are
unaccounted for by the PSF model (Anderson et al.\ 2008).  In an
effort to remove the latter two effects, we adopted 
the procedure described in Milone et al.\ (2011), which we briefly 
summarise in the following.

\begin{figure*}[ht!]
   \centering
   \includegraphics[width=13 cm]{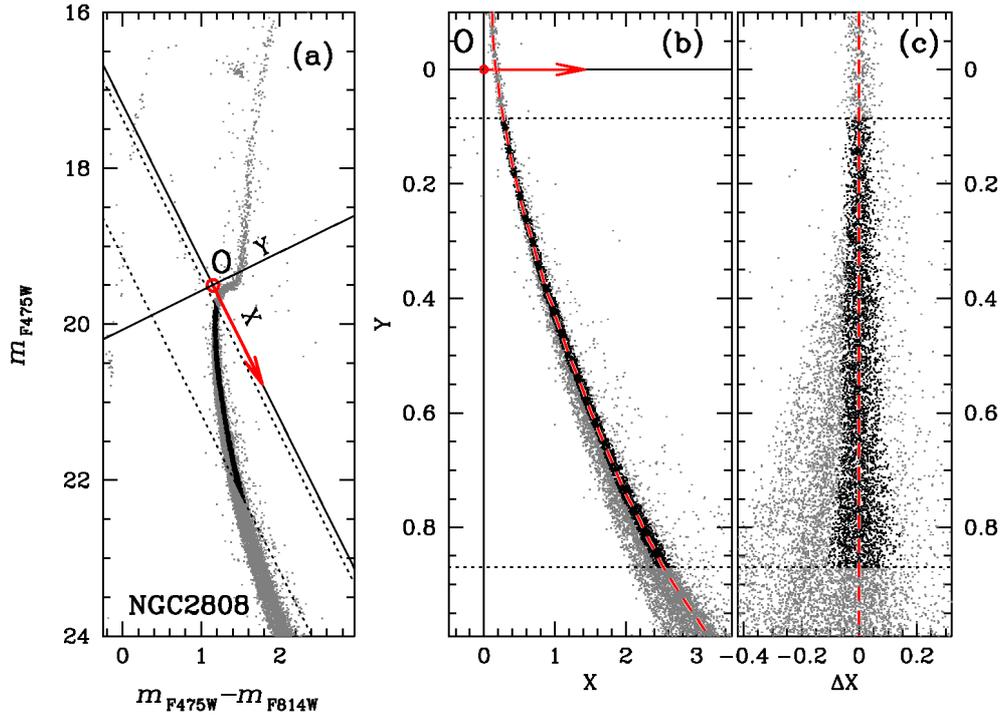}
   \caption{
 \textit{Panel a} CMD of NGC 2808; the red arrow
	indicates the direction of reddening.  The continuous lines
	are the axes (${\mathcal X}$ and ${\mathcal Y}$) of the
	reference frame introduced during the procedure for the
	measurements of reddening variations.  The location of stars
	in this reference frame is shown in \textit {panel b}, where we
	draw the fiducial line of the MS as a dashed red line.
	Reference stars are indicated with black dots and are located
	between the dotted lines.  The rectified ${\mathcal Y}$ vs.\
	$\Delta {\mathcal X}$ diagram is plotted in the \textit {panel
	c}.  }
         \label{procred}
   \end{figure*}

For simplicity, we defined the photometric reference frame shown in
Fig.~\ref{procred}, where the abscissa is parallel to the reddening
line.  To do this, we first arbitrarily defined a point ({\rm O})
near the MS turn-off in the CMD of panel a.  We then translated the
CMD such that the origin of the new reference frame corresponds to
{\rm O}. Finally, we rotated the CMD counterclockwise by an
angle
\begin{center}
 $ \theta={\rm arctan} \frac{A_{\rm F475W}}{A_{\rm F475W}-A_{\rm F814W}} $,
\end{center}
as shown in Fig.~\ref{procred}b.  
For clarity, in the following, we
name the abscissa and the ordinate of the rotated reference frame, 
 ${\mathcal X}$ and ${\mathcal Y}$.  The two quantities $A_{\rm
F475W}$ and $A_{\rm F814W}$ are the absorption coefficients in the
F475W and F814W ACS bands that, according to Bedin et al.\ (2005),
correspond to the average reddening of NGC 2808 [$E(B-V)=0.22$ Harris
1996, 2003, $A_{\rm F475W}=0.80$, $A_{\rm F814W}=0.40$].

At this point, we adopted an iterative procedure to determine a locally
valid estimate of $A_{\rm F475W}$ and $A_{\rm F814W}$ that involves
the following four steps:
\begin{enumerate}
\item
We used the rMS stars to generate the red fiducial line shown in
Fig.~\ref{procred}b.  We divided the sample of these reference stars
into bins of 0.2 mag in ${\mathcal Y}$.  For each bin, we calculated the
median ${\mathcal X}$ that had been associated with the median
${\mathcal Y}$ of the stars in the bin.  The fiducial 
was then derived
by fitting these median points with a cubic spline.  Here, it is
important to emphasise that the use of the median allows us to
minimise the influence of the outliers such as contamination by bMS, any
mMS stars left in the sample, binary stars, or stars with poor
photometry.
\item
For each star, we calculated the distance from the fiducial line along
the reddening direction ($\Delta {\mathcal X}$).  In the right panel
of Fig.~\ref{procred}, we plot ${\mathcal Y}$ versus $\Delta {\mathcal
X}$.
\item
We selected the sample of stars located in the regions of the CMD
where the reddening and the fiducial line define a wide angle such that
the shift in colour and magnitude due to differential reddening can be
more easily separated from the random shift due to photometric errors.
These stars are used as reference stars to estimate reddening
variations associated with each star in the CMD and are marked in
Fig.~\ref{procred} as heavy black points.
\item
The basic idea of our procedure, which is applied to each star
(target) individually, is that differential reddening and spatially
dependent PSF errors both have the effect that the stars in a local
region are all shifted by the same amount to either the blue or red and in
magnitude, relative to the fiducial sequence.  The adopted size of
the comparison region is a compromise between two competing needs. On
one hand, we wish to use the smallest possible spatial cells, so that
the systematic offset between the 'abscissa' and the fiducial
ridgeline will be measured as accurately as possible for each star's
particular location.  On the other hand, we wish to use as many stars
as possible, to reduce the error in the determination of the
correction factor.

As a compromise, for each star we selected the nearest 60 reference
stars, determined the median $\Delta {\mathcal X}$, and assumed this to
be the reddening correction for that star.
In calculating the differential reddening suffered by a reference
star, we naturally
excluded the star itself in the computation of the median
$\Delta {\mathcal X}$.
\end{enumerate}
After the median $\Delta {\mathcal X}$ had been subtracted from the
${\mathcal X}$ value of each star in the rotated CMD, we obtained an
improved CMD, which was then used to derive a more accurate selection
of the sample of rMS reference stars and derive a more precise
fiducial line.

After step 4, we had a newly corrected CMD.  We re-ran the procedure
to verify whether the fiducial sequence needed to be slightly changed in
response to the adjustments made.  The procedure converged after five
iterations.  Finally, the corrected ${\mathcal X}$ and ${\mathcal Y}$
were converted 
back to $m_{\rm F475W}$ and $m_{\rm F814W}$ magnitudes.
%
%

\begin{figure*}[ht!]
   \centering
   \includegraphics[width=13 cm]{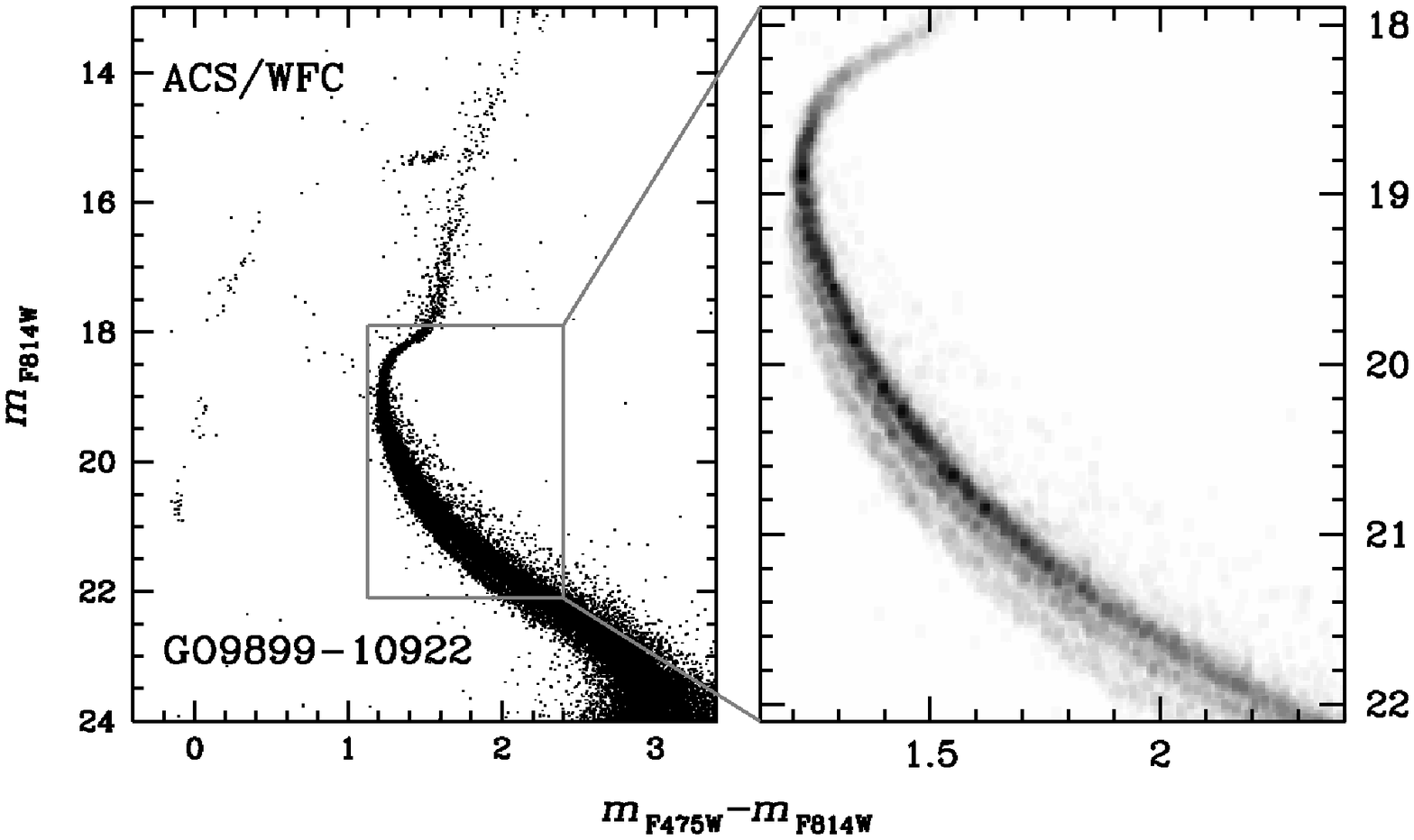}
      \caption{\textit{Left panel:} CMD from ACS/WFC photometry
        (GO9899 and GO10922). \textit{Right panel:} Hess diagram for
        the MS region (indicated with a grey box on the left) studied
        in this paper.}
         \label{CMD10992}
   \end{figure*}

The result of the correction is that all the sequences of the CMD are
 more sharply defined.  A zoom of the corrected ${\it m}_{\rm F814W}$ versus
${\it m}_{\rm F475W}-{\it m}_{\rm F814W}$ plot around the SGB and the
upper-MS region is shown in the right panel of Fig.~\ref{COMRED}, where
the effects of differential reddening in the original CMD are
particularly evident.  The whole CMD corrected for differential
reddening is plotted in the left panel of Fig.~\ref{CMD10992}. The
Hess diagram shown in the right panel highlights the triple MS of NGC
2808, which is the main subject of the present study.
%
\section{Artificial stars}
\label{AS}
The artificial-star (AS) experiments were carried out using the
procedure and the software described in Anderson et al.\ (2008).  We
first produced an input list with about $5 \times 10^{5}$ stars
distributed across the entire WFC field of view. This list includes the 
coordinates of the stars in the reference frame and the magnitudes in
the F475W and F814W bands.  We placed artificial stars along the three
MSs assuming a flat LF in the F814W band. Stellar positions were
determined 
according to the overall cluster radial distribution by following the
recipe of Milone et al.\ (2009).

For each object in the input list, the program described in Anderson et
al.\ (2008) adds a star into each image with the appropriate position
and flux, and measures it with the same procedure as for real stars.  The
AS tests are performed for one star at a time such that 
artificial stars never interfere with each other.
We considered an AS as
recovered when the input and the output position differed by less than
0.5 pixel and the input and output magnitude by less than 0.75 mag.

Artificial stars  
are used for several purposes:
(i) to measure the completeness level of our photometry,
(ii) to estimate the fraction of chance-superposition binaries, and
(iii) to determine the LF of the three MSs.
\subsection{Completeness}
Completeness was calculated by following the procedure described 
in Milone et al.\ (2009), which accounts for both crowding conditions
and stellar brightness. To account for stars of the
same luminosity, but belonging to different MSs, having different
colours, hence different completeness, we separately estimated the
completenesses of the red, the middle, and the blue MS.

We, briefly, divided the ACS/WFC field into five annuli centred on the
cluster centre and analysed AS results in seven magnitude intervals.
We calculated the average completeness corresponding to each of these
$5\times7$ grid points, as the ratio of the recovered to input
artificial stars within that bin, and estimated the completeness for
any star at any position within the cluster by interpolating 
among the grid points.

\begin{figure*}[ht!]
   \centering
   \includegraphics[width=13.5 cm]{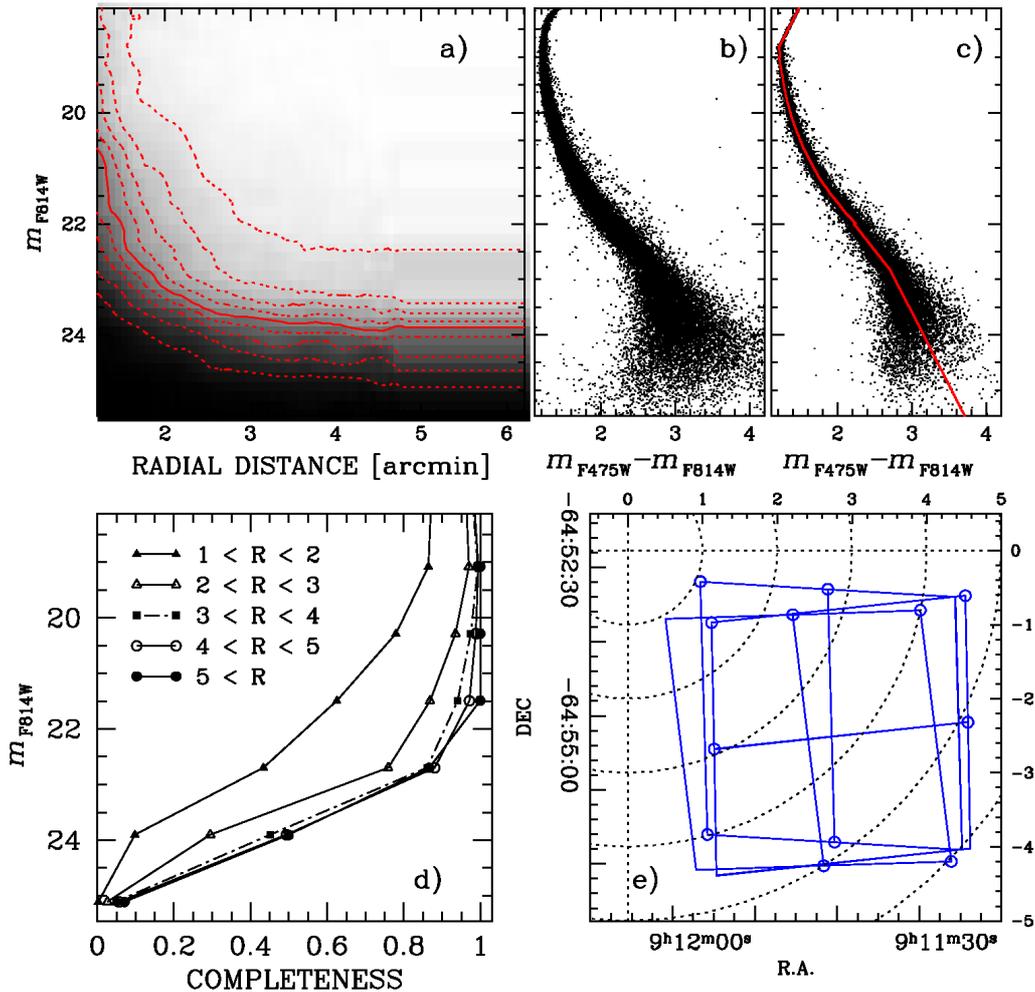}
   \caption{\textit{Panel a}: Completeness contours in the ${\it
   m}_{\rm F814W}$ versus radial distance 
   plane. The continuous line marks the completeness level of 0.50,
   while dotted lines indicate differences in completeness in steps of
   0.10 units.  The observed CMD and the CMD for AS (as added in red,
   as recovered in black) are plotted in \textit{panels b} and
   \textit{c}, respectively.  \textit{Panel d} shows the completeness
   for rMS stars as a function of ${\it m}_{\rm F814W}$ in the five
   annuli into which we divided the field of view. The footprints of our data are 
   shown in \textit{panel e}. [See text for details.]  }
   \label{completeness}
   \end{figure*}

The result of this procedure is illustrated in Fig.~\ref{completeness}
for the case of the rMS.  Panel a shows the completeness contours in
the observational plane ${\it m}_{\rm F814W}$ versus radial distance
from the centre of the cluster.
The grey levels indicate the completeness of our sample with white and
black colour codes corresponding to a completeness of 1 and 0, 
respectively.  The continuous line corresponds to a completeness level
of 0.50, while dotted lines indicate differences in completeness of
0.10.  The observed CMD is plotted in panel b and the CMD of
artificial stars in panel c. In the latter, we represented with red
and black colour codes the input ASs and the observed ones,
respectively.  Panel d shows the completeness for rMS stars as a
function of ${\it m}_{\rm F814W}$ in the five annuli into which 
we divided the
field of view. The footprints of our data are represented in panel e,
with the boundaries of each annulus marked by dotted circles.

%

\section{Fiducial lines and widths of the three MSs}
\label{sec:fiducial}
We now describe how we determined the fiducial lines for the three MSs and
estimated their color dispersions as a function of magnitude. 
These
results provide crucial information to help us characterise the stellar
populations of NGC 2808 and are useful tools for measuring the LFs and
the fraction of binaries in this GC 
(see Sect.~\ref{LUFU}).

    \begin{figure*}[ht!]
    \centering
    \includegraphics[width=7.25 cm]{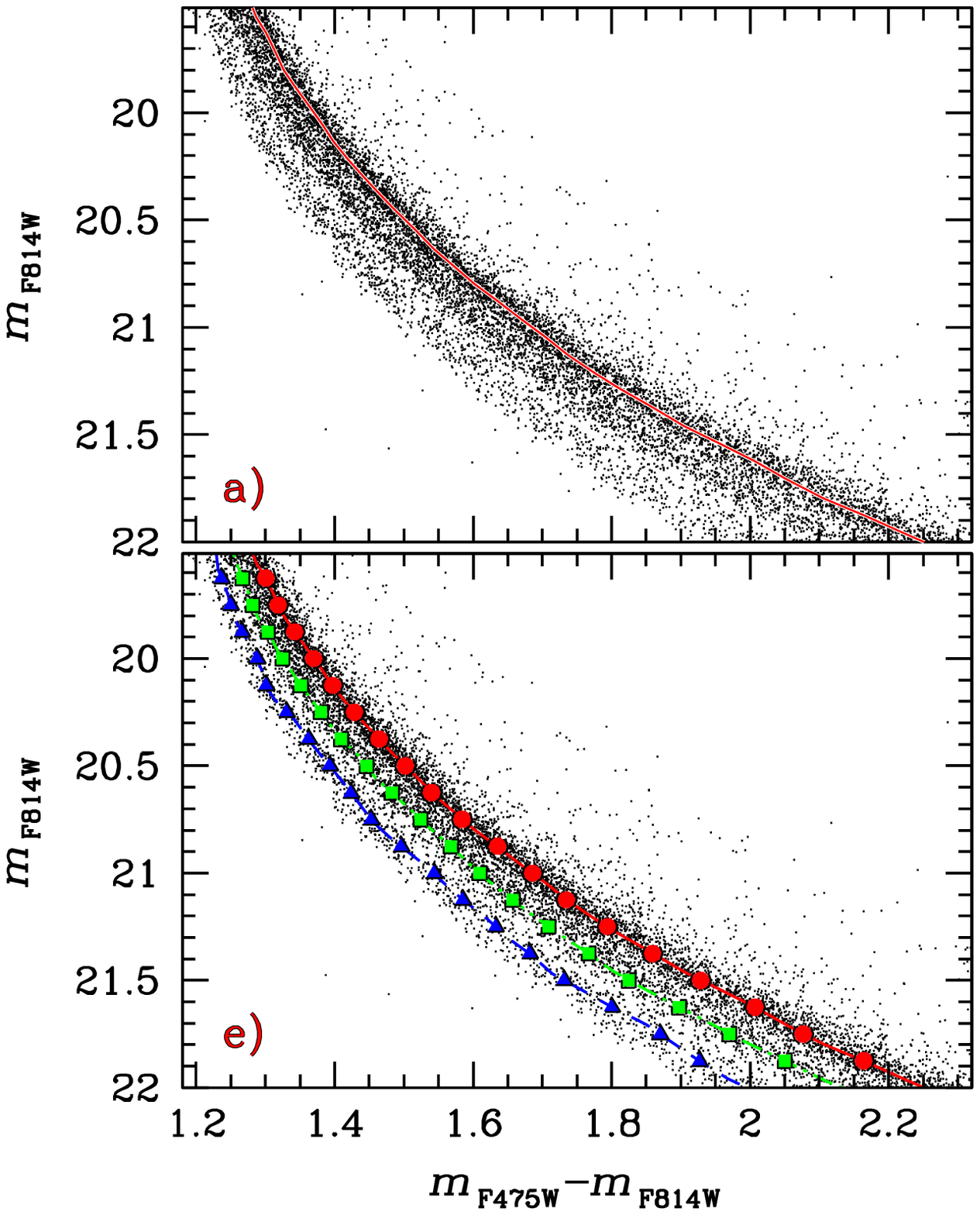}
    \includegraphics[width=9.75 cm]{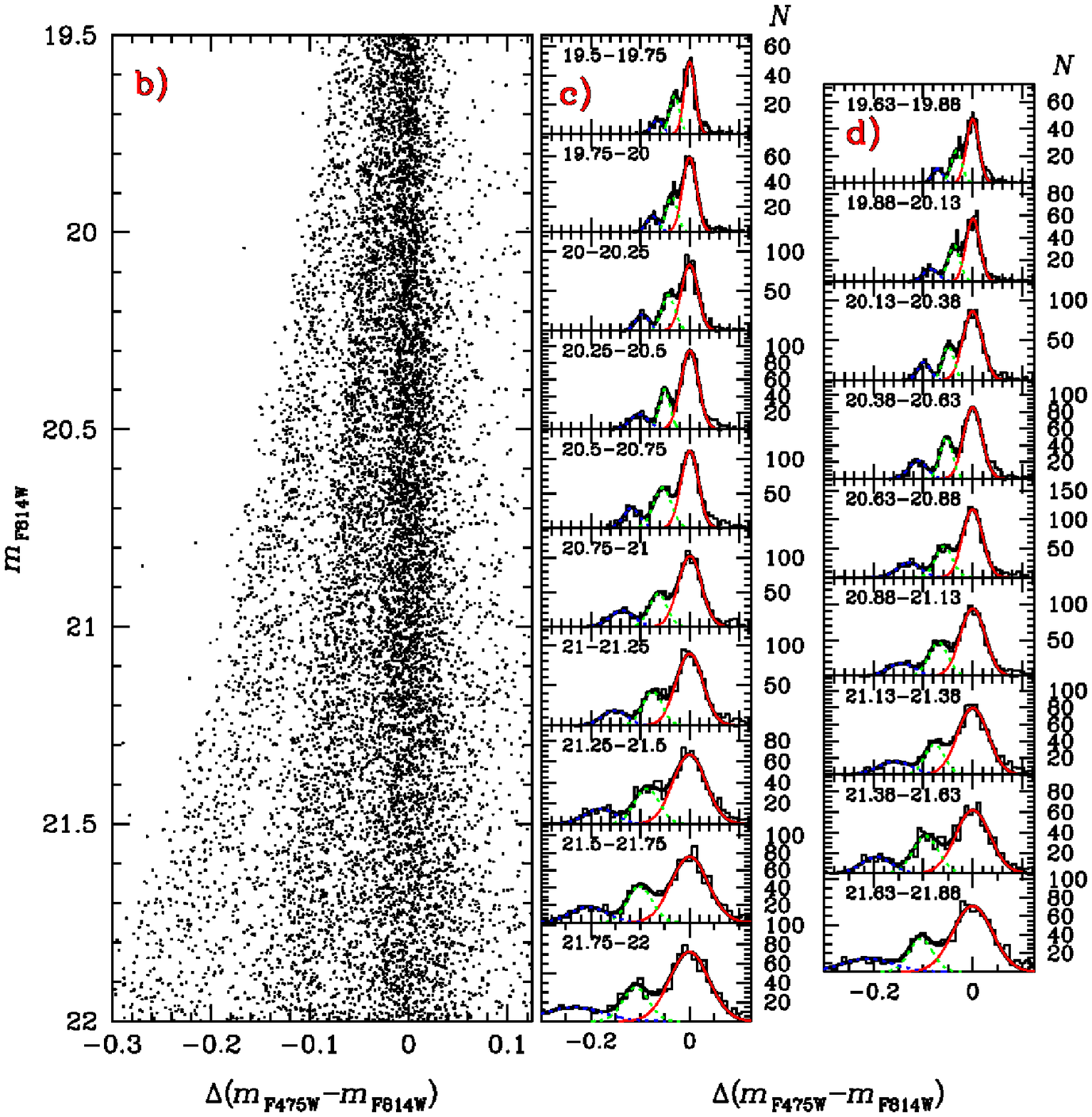}
       \caption{
        This figure illustrates the procedure to determine the rMS
 	fiducial line.  A region of the CMD focused on the triple MS
 	of NGC 2808 is shown in \textit{panel a}. The continuous line
 	is the first-guess fiducial for the rMS, drawn by
 	hand. \textit{panel b}. The same CMD, after subtracting from
 	the colour of each star the corresponding colour on the
 	fiducial at the same magnitude. \textit{Right panels c} and
 	\textit{d} show the colour distribution of stars in the 0.25
 	F814W magnitude intervals quoted in the inset. The continuous
 	black lines are least squares fits of a sum of the three
 	Gaussians that are represented as blue, green, and red
 	lines. The $\Delta(m_{\rm F475W}-m_{\rm F814W})$ values
 	corresponding to the peak of the red Gaussian and used to
 	derive the red MS fiducial line are superimposed on the CMD of
 	\textit{panel e} (red filled circles).  For completeness, in
 	the latter panel we also show the fiducials for the bMS (blue
 	triangles) and the middle MS (green squares).}
       \label{fiducials}
    \end{figure*}

\begin{table}[ht!]
\begin{center}
\caption{
Magnitudes ($ \textit{mag} = m_{\rm F814W}$), 
colours ($\textit{col} =  m_{\rm F475W}-{\it m}_{\rm  F814W}$), and
colour spreads of 
the fiducial lines for the three MSs shown in Fig.~\ref{fiducials}.
}
\begin{tabular}{ccccccc}
\hline
\hline
\textit{mag} & $col_{\rm rMS}$  & $col_{\rm mMS}$  & $col_{\rm bMS}$ & $\sigma_{\rm rMS}$  & $\sigma_{\rm mMS}$  & $\sigma_{\rm bMS}$\\
 \hline
   19.625 &   1.300 &   1.270 &   1.233 & 0.011 & 0.009 & 0.013\\
   19.750 &   1.318 &   1.287 &   1.248 & 0.012 & 0.010 & 0.012\\
   19.875 &   1.343 &   1.310 &   1.266 & 0.014 & 0.011 & 0.011\\
   20.000 &   1.369 &   1.330 &   1.286 & 0.014 & 0.009 & 0.014\\
   20.125 &   1.397 &   1.355 &   1.303 & 0.017 & 0.013 & 0.013\\
   20.250 &   1.430 &   1.385 &   1.330 & 0.019 & 0.012 & 0.013\\
   20.375 &   1.466 &   1.415 &   1.356 & 0.018 & 0.011 & 0.015\\
   20.500 &   1.504 &   1.451 &   1.390 & 0.018 & 0.012 & 0.019\\
   20.625 &   1.541 &   1.486 &   1.422 & 0.017 & 0.015 & 0.016\\
   20.750 &   1.588 &   1.528 &   1.453 & 0.020 & 0.017 & 0.020\\
   20.875 &   1.638 &   1.573 &   1.496 & 0.023 & 0.017 & 0.024\\
   21.000 &   1.687 &   1.617 &   1.540 & 0.025 & 0.019 & 0.025\\
   21.125 &   1.736 &   1.668 &   1.578 & 0.027 & 0.019 & 0.028\\
   21.250 &   1.796 &   1.720 &   1.631 & 0.031 & 0.021 & 0.028\\
   21.375 &   1.863 &   1.778 &   1.681 & 0.033 & 0.025 & 0.030\\
   21.500 &   1.930 &   1.835 &   1.728 & 0.035 & 0.023 & 0.029\\
   21.625 &   2.008 &   1.908 &   1.793 & 0.037 & 0.024 & 0.037\\
   21.750 &   2.078 &   1.977 &   1.855 & 0.042 & 0.027 & 0.035\\
   21.875 &   2.163 &   2.052 &   1.920 & 0.042 & 0.034 & 0.046\\
\hline
\hline
\end{tabular}
\end{center} 
\label{tablefiducials}
\end{table}

The fiducial sequences and the color dispersions were determined for
each MS separately using the procedure illustrated in
Fig.~\ref{fiducials}, described as follows for the case of the rMS.
We limited our analysis to MS stars with $19.5<{\it m}_{\rm
F814W}<22$, as the three MSs can be more easily distinguished in this magnitude
interval.  A zoom of the ${\it m}_{\rm F814W}$ vs.\ ${\it m}_{\rm
F475W}-{\it m}_{\rm F814W}$ CMD of Fig.~\ref{CMD10992} is shown in
panel a, where the red line is a first-guess fiducial line for the rMS
drawn by hand.  In panel b, we subtracted from the color of each
star the colour of the rMS fiducial line at the same magnitude of the
star.  The following two panels show the colour distribution of the
stars in panel b.  Specifically, in panels c we plotted the colour
distribution in ten intervals of 0.25 F814W magnitudes from ${\it
m}_{\rm F814W}=19.5$ to ${\it m}_{\rm F814W}=22.0$, while panel d
shows the color distribution in nine $m_{\rm F814W}$ bins between
19.75 and 21.75.

As widely discussed in Piotto et al.\ (2007), the histogram
distributions clearly show three distinct peaks, which we 
fitted using a least squares method 
with the sum of three Gaussians represented as a continuous
black line. The single Gaussian components that fit each MS are
plotted with blue, green, and red lines.  The $\Delta(m_{\rm
F475W}-m_{\rm F814W})$ values corresponding to the peak of the red
Gaussian are plotted as red-filled circles in the CMD of \textit{panel
e}.  The rMS fiducial line was obtained by fitting these red
squares by means of a spline and is represented as a continuous red
line.

Note that both the bMS and the mMS are not vertical in the $m_{\rm
F814W}$ vs.\ $\Delta(m_{\rm F475W}-m_{\rm F814W})$ diagram of
\textit{panel b}.  As a consequence of this, the dispersions of the
blue and green Gaussian are not used to estimate the color spread and
the fiducials of these MSs. To determine these quantities, we instead followed a
procedure similar to the one just described for the rMS, but in these
cases we subtracted from the observed colour of each star, the
colour of the mMS (or the bMS) at the corresponding F814W magnitude.
For completeness, the fiducials of the mMS (in green) and bMS (in blue)
are also plotted in \textit{panel e}, while the $m_{\rm F475W}-m_{\rm
F814W}$ colours and the $m_{\rm F814W}$ magnitudes of the three
fiducial lines are listed in Table~1.
   \begin{figure}[ht!]
   \centering
   \includegraphics[width=8.5 cm]{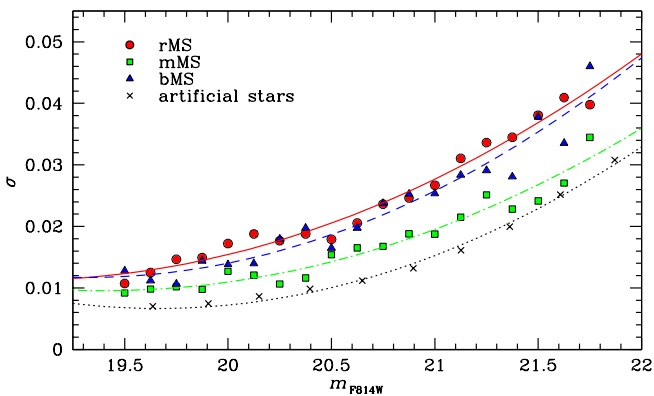}
   \caption{Observed color dispersion of the bMS (blue triangles), mMS
     (green squares), rMS (red circles), and AS (black crosses), as a
     function of the F814W magnitude. Continuous, dashed-dotted,
     dashed, and dotted lines are the best-fit parabolas for the
     rMS, mMS, bMS, and the rMS AS, respectively.}
   \label{sigfmas}
   \end{figure}

The dispersions obtained for the bMS ($\sigma_{\rm bMS}$), mMS
($\sigma_{\rm mMS}$), and rMS ($\sigma_{\rm rMS}$), and the dispersion
of artificial stars ($\sigma_{\rm AS}$) are plotted in
Fig.~\ref{sigfmas} as a function of $m_{\rm F814W}$. The continuous,
dashed-dotted, dashed, and dotted lines are the best-fit
second-order polynomials for the three MSs (${\it P}_{\rm bMS, mMS,
rMS}$) and for the simulated CMD ($P_{\rm AS}$).

As expected, the MS spread for the artificial stars is slightly
smaller than the colour spread of real stars. This is the consequence
of an important limitation of artificial-star tests, and is widely
discussed in Anderson et al.\ (2008) and Milone et al.\ (2009) for the
case of ACS/WFC images.  While artificial stars are measured by using
the same PSFs used to generate them, for real stars we are unable to
obtain a perfect PSF model. The PSF for a real star is constructed to
reproduce as well as possible the profile of the real stars, but we
cannot avoid having some errors in the PSF model; our real-star
measurements will suffer from these errors, 
but our artificial-star
measurements will not. We note that this additional
dispersion for real stars does not allow us to exclude the effect of
the intrinsic 
broadening of the MSs, which could be caused by an intrinsic dispersion
in $Z$, $Y$, or a combination of the two (see e.\ g.\ Marino et al.\
2009 and Anderson et al.\ 2009).  In this context, 
we note that while the measured $\sigma_{\rm bMS}$ values do not
differ significantly from $\sigma_{\rm rMS}$, the mMS does have
a marginally smaller dispersion over the whole range of magnitude we
analysed.  It is tempting to conclude that the different colour width
of the mMS could be due to a smaller intrinsic spread than those of
both bMS and rMS.  However, our data do not allow us to conclude that
this difference is significant.

So that we can compare the observed and simulated 
          MSs we added an additional error component into the 
          artificial-star colors. To do this, we added to the
          colour of each bMS 
(mMS, rMS) artificial star an error randomly extracted from a Gaussian
distribution with a dispersion $P_{\rm DIF}=\sqrt{P_{\rm bMS (mMS,
rMS)}^{2}-P_{\rm AS}^{2}}$ (see Milone et al.\ 2009 for more details).

\section{The luminosity function of the three MSs}
\label{LUFU}
We present the recipe used to determine both the
fraction of binaries and the LFs of the three stellar populations of
NGC 2808.  Specifically, Sect.~\ref{sub:iso} describes the
stellar models used to determine the mass-luminosity relations
involved in the measurement of the binary fraction and estimate
the MFs.  Sect.~\ref{BIN} analyses the behaviour of binaries in the
CMD of NGC 2808, while in Sect.~\ref{sub:LF} we describe in detail
the procedure used to measure the LFs and the fraction of binaries
and discuss the results.

\subsection{Isochrones}
\label{sub:iso}
To convert the LFs into MFs, we used 
the mass-luminosity relation from the stellar models in Dotter et al.\
(2007), and an independent set of models specifically computed 
 for this 
project.  At the low-mass end, these additional stellar models
correspond to the very low-mass (VLM) range (${\mathcal
M}<0.5{\mathcal M}_{\rm \odot}$) and rely on the equation of state and
the boundary conditions in Cassisi et al.\ (2000). For more massive
stars, we adopted the same physical scenario adopted by Pietrinferni
et al.\ (2004, the BaSTI archive\footnote{\texttt{http://albione.oa-teramo.inaf.it}}). 
Considerable care was devoted to
performing an accurate match ---avoiding any discontinuity that could
affect the resulting mass-luminosity relation--- between stellar models
in the VLM regime and more massive structures. The stellar models were
 used to compute various sets of isochrones for a metallicity
suitable for NGC 2808 and various He abundances.  The whole set of
models (hereafter BaSTI models) were transformed from the
theoretical plane to the ACS observational one by means of
theoretical color transformations and bolometric corrections for the
standard Johnson bandpasses provided by Allard et al.\ (1997) and,
then applying the equations provided by Sirianni et al.\ (2005) to
convert the ground-based Johnson magnitudes to the \textit{HST\,} ACS filters.

Since spectroscopic studies have demonstrated that NGC 2808 stars have
homogeneous iron contents, the only way to fit the three MSs is to
assume that the rMS has a canonical He content, while the mMS and the bMS
are matched by He enriched isochrones (Piotto et al.\ 2007).  We were
able to fit the rMS with an isochrone with $Y=0.248$ and the closest
matches between the mMS and bMS were obtained by assuming that
$Y\sim0.31$-32 and $Y\sim0.38$, respectively.

\begin{figure}[ht!]
   \centering
   \includegraphics[width=8.5 cm]{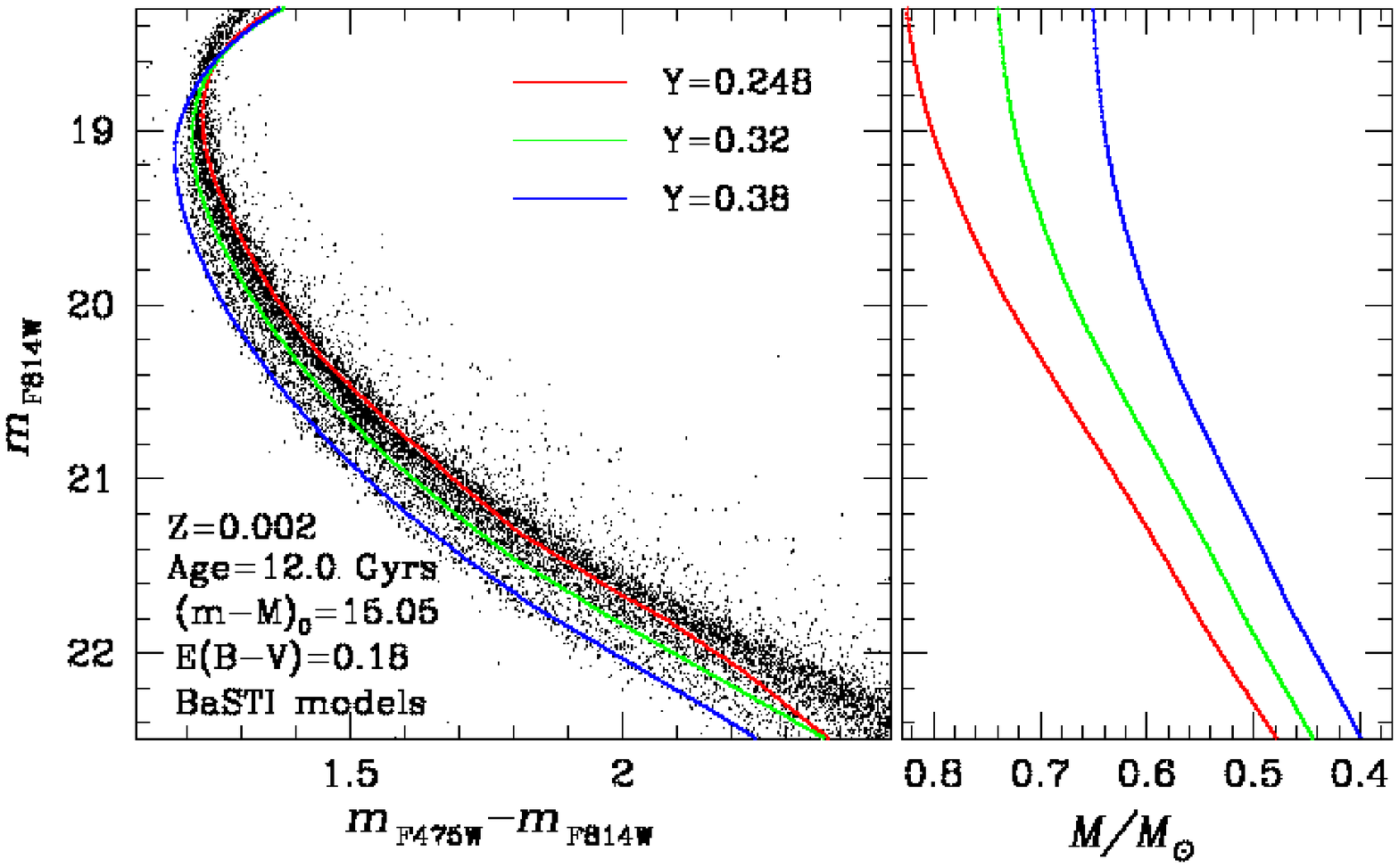}
   \includegraphics[width=8.5 cm]{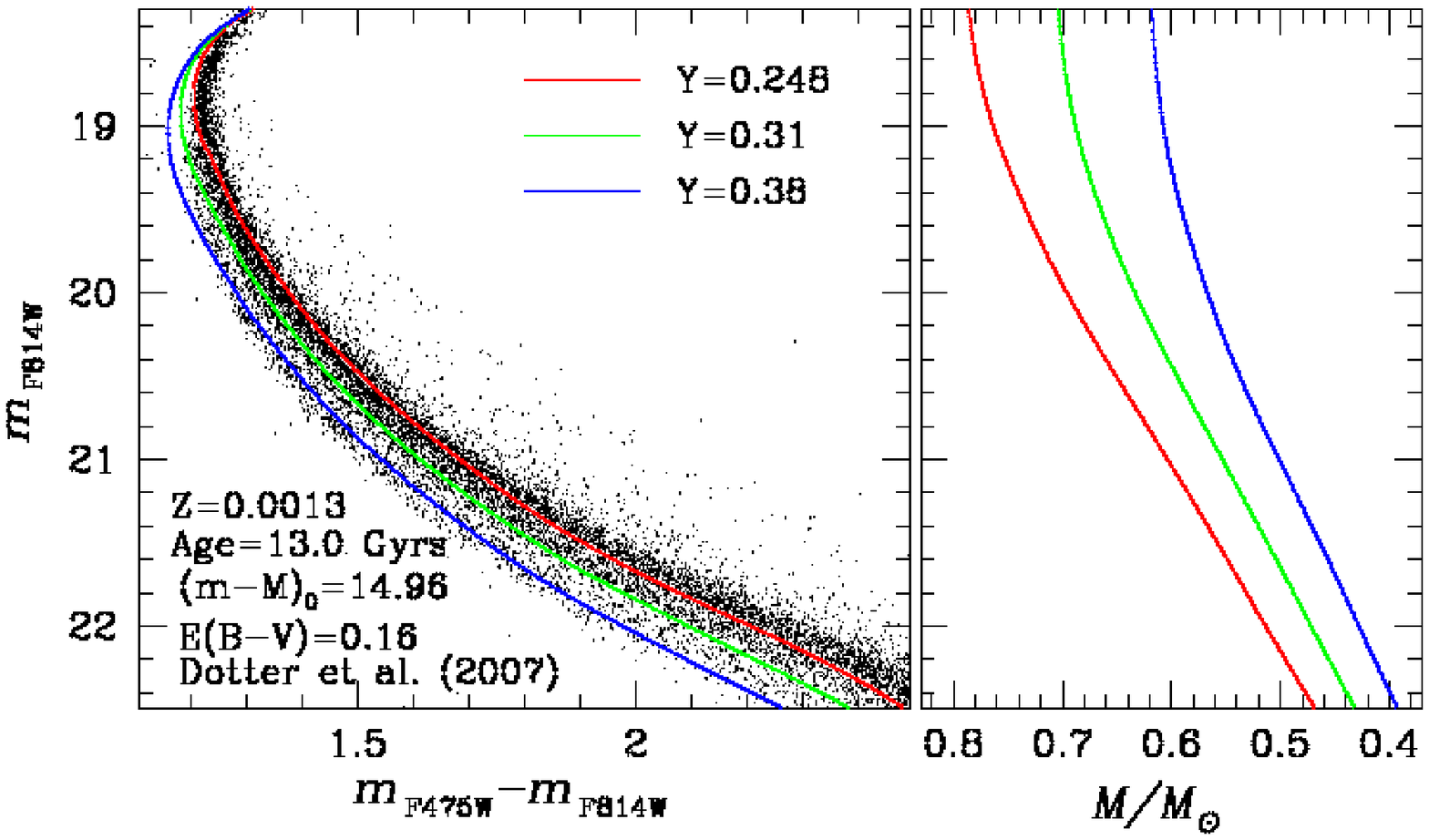}
   \caption{\textit{Left panels}. Best-fit isochrones from
        Pietrinferni et al.\ (2004) (\textit{top}) and Dotter et al.\
        (2007) (\textit{bottom}) superimposed on the CMD of
        Fig.~\ref{CMD10992}.  Best-fit parameters are listed in
        the inset.  \textit{Right panels} show the luminosity in the
        F814W band as a function of the stellar mass. }
   \label{iso}
   \end{figure}

The comparison between the observed CMD and the isochrones is shown in
the left panels of Fig.~\ref{iso} for the Dotter et al.\ (2007)
isochrones (bottom panel) and for the BaSTI models (upper panel).  The
values of metallicity, $Y$, distance modulus, and foreground reddening
that provide the closest match are provided in the inset of
Fig.~\ref{iso} and are listed in Table~2.

The three stellar populations should have almost the same age but
stars along the bMS and the mMS are less massive than rMS stars of
the same luminosity.  Right panels show the relation between ${\it
m}_{\rm F814W}$ and stellar mass predicted by theoretical models for
stars of the triple MS.

From Fig.~\ref{iso}, it appears that both sets of isochrones do not
accurately reproduce the turn-off portion of the CMD. The fit could be
slightly improved by adopting a different choice of cluster age,
but in any case some disagreement would be still present as a
consequence ---probably--- of some residual drawback in the available
color-effective temperature relation and bolometric correction scale. 
Nevertheless,
we wish to emphasize that this mismatch between
 theory and observations in the turn-off region does not affect
our present investigation, since we measure the MF far below the mass
where evolutionary effects are significant.

%
%

\begin{table*}[ht!]
\label{parametri}
\begin{center}
\caption{ Basic properties of NGC 2808 inferred from the isochrones
  used in the paper.  }
\begin{tabular}{cccccc}
\hline
\hline
Model          &
  $Z$ &
  Age          &
  ${\rm (m-M)_{0}}$ &
  {\it E(B-V)}        &
  $Y$(rMS - mMS - bMS) \\
 \hline
BaSTI                    & 0.0020  & 12.0 Gyrs  & 15.05   & 0.18  &
0.248 - 0.32 - 0.38\\
Dotter et al.\ (2007)    & 0.0013  & 13.0 Gyrs  & 14.96   & 0.16  &
0.248 - 0.31 - 0.38\\
\hline
\hline
\end{tabular}
\end{center}
\end{table*}

\subsection{Photometric MS-MS binaries}
\label{BIN}
At the distance of NGC 2808 ($d\sim9.6$ Kpc, Harris\ 1996, update
2003), binary stars are completely unresolved, and the light from
both stars combines to form a single point-like source.
Specifically, if we indicate with $m_{1}$ and $m_{2}$ the magnitudes, and
$F_{1}$ and $F_{2}$ the fluxes of the two components, the binary will
appear as a single object with a magnitude
\begin{center}
$m_{\rm bin}=m_{1}-2.5~\log{(1+\frac{F_{1}}{F_{2}})}$.
\end{center}

For a simple stellar population, the luminosity of binary
systems formed by two MS stars depends on the mass ratio
($q=\mathcal{M}_{2}/\mathcal{M}_{1}, \mathcal{M}_{2} <
\mathcal{M}_{1}$) of the two components as stellar masses
($\mathcal{M}_{1}$, and $\mathcal{M}_{2}$) are 
directly related to
the luminosity.  The equal-mass binaries form a sequence
that is parallel to the MS, but about 0.75 magnitudes brighter.  When the
masses of the two components are different, the binary will appear
redder and brighter than the primary and populate a CMD region on the
red side of the MS ridge line (MSRL) and below the equal-mass binary
line.  As a consequence of this, the binaries formed by stars of
similar mass can be easily discerned photometrically from the single
MS stars as they have large offsets from the 
MSRL, whereas in the CMD
it is hard to distinguish binary systems formed by stars that
strongly differ in mass because of their small colour offset from the
MSRL.

To calculate the total fraction of binaries in NGC 2808, we
follow an approach adopted by several authors (e.\ g.\
Rubenstein \& Bailyn \ 1997, Zhao \& Bailyn \ 2005, Sollima et al.\
2007, Milone et al.\ 2010, 2011) that consists of simulating a binary
population that
follows a given distribution of mass ratios [{\it
f}({\it q})] and in comparing the observed and simulated CMDs.

The adopted {\it f}({\it q}) is crucial for a correct measurement of
the binary fraction. To date, the only constraint on the shape of {\it
f}({\it q}) in GCs comes from the work of Milone et al.\ (2011), who
studied the main properties of the population of MS-MS binaries in 59
Galactic GCs and found that the mass-ratio distribution is generally
flat for $q>0.5$.  Unfortunately, these authors were unable to place any
constraints on mass ratios smaller than $q=0.5$.  For simplicity, in
this work, we extrapolate the flat mass-ratio distribution all
the way down to $q=0$.

In Appendix A, we derive the fraction of binaries and the MFs of
the three MSs by using a different mass-ratio distribution. We 
show that the use of a different {\it f}({\it q}) does not change the
main conclusions of this paper.

A further complication, in the case of NGC 2808, 
is the triple stellar population.  Until now, there have been
neither observational nor theoretical constraints on the
origin of the components of binary systems in GCs with multiple MSs.
In particular, we do not know whether the components of a binary
system belong preferentially to the same stellar population or
originate from different MSs.

   \begin{figure*}[ht!]
   \centering
   \includegraphics[width=12.5 cm]{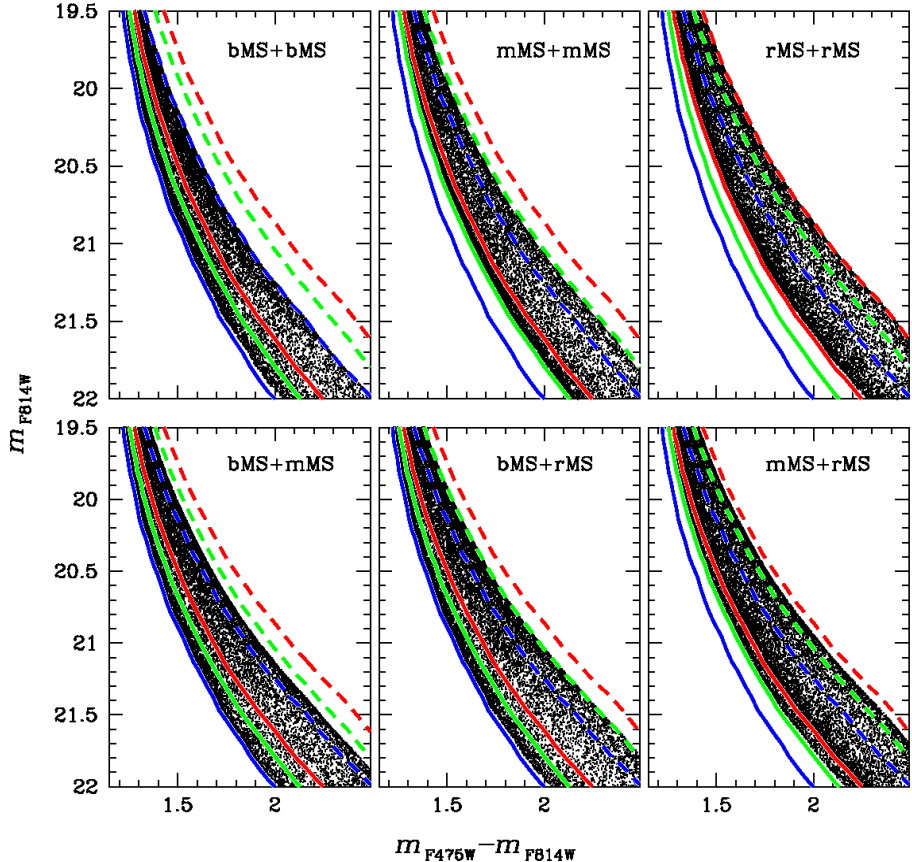}
      \caption{\textit{Upper panels:} Simulated CMDs for MS-MS
        binaries formed by pairs of bMS \textit{(left)}, mMS
        \textit{(middle)}, and rMS stars
        \textit{(right)}. \textit{Bottom panels:} binaries composed by
        bMS and mMS \textit{(left)}, bMS and rMS \textit{(middle)},
        and mMS and rMS stars \textit{(right)}. Continuous lines are
        the empirical MSRLs for the three MSs while sequences of equal
        mass bMS, mMS, and rMS binaries are marked with dashed lines.
        }
         \label{binarie}
   \end{figure*}
%

In the upper panels of Fig.~\ref{binarie}, we used our empirical MSRLs
and the mass-luminosity relations of Pietrinferni et al.\ (2004) to
generate sequences of MS-MS binary systems formed by pairs of stars
belonging to the same population (i.\ e.\ pairs of bMS+bMS, mMS+mMS,
and rMS+rMS stars).  The bottom panels show the CMDs for binaries made
of stars from different MSs.  (This is not inconceivable; GCs are dynamically
evolving objects and binaries can be destroyed and re-created over
a Hubble time.)
The
MSRLs of the bMS, mMS, and rMS have been represented with blue, green,
and red continuous lines and we have marked with dashed lines the sequences
of equal mass bMS, mMS, and rMS binaries. In all cases, we assumed
a flat mass-ratio distribution.

While the determination of the frequency of each of these groups of
binaries is beyond the scope of this paper, it is clear that the
results of this paper may depend in some way on the different scenario
we adopt.  To account for this, we determine the binary fraction
and the LFs by assuming two extreme scenarios.  In 
Sect.~\ref{sub:LF}, we describe in detail the adopted procedure
considering only one of these extreme scenarios which assumes that
each component of a binary system has the same probability of belonging
to any MS.

In Appendix B, the other extreme scenario is considered, i.\ e.\
that both the components of all binary systems belong to the same 
MS.  We demonstrate that the two extreme scenarios predict differences
in the binary fraction that are less than 0.01, hence that the
obtained MFs do not depend significantly  on the adopted scenario.

\subsection{Luminosity functions }
\label{sub:LF}
We describe the iterative procedure used to derive
both the LFs of the three MSs and the total binary fraction ($f^{\rm
BIN}$) in NGC 2808.

   \begin{figure*}[ht!]
   \centering
   \includegraphics[width=12.5 cm]{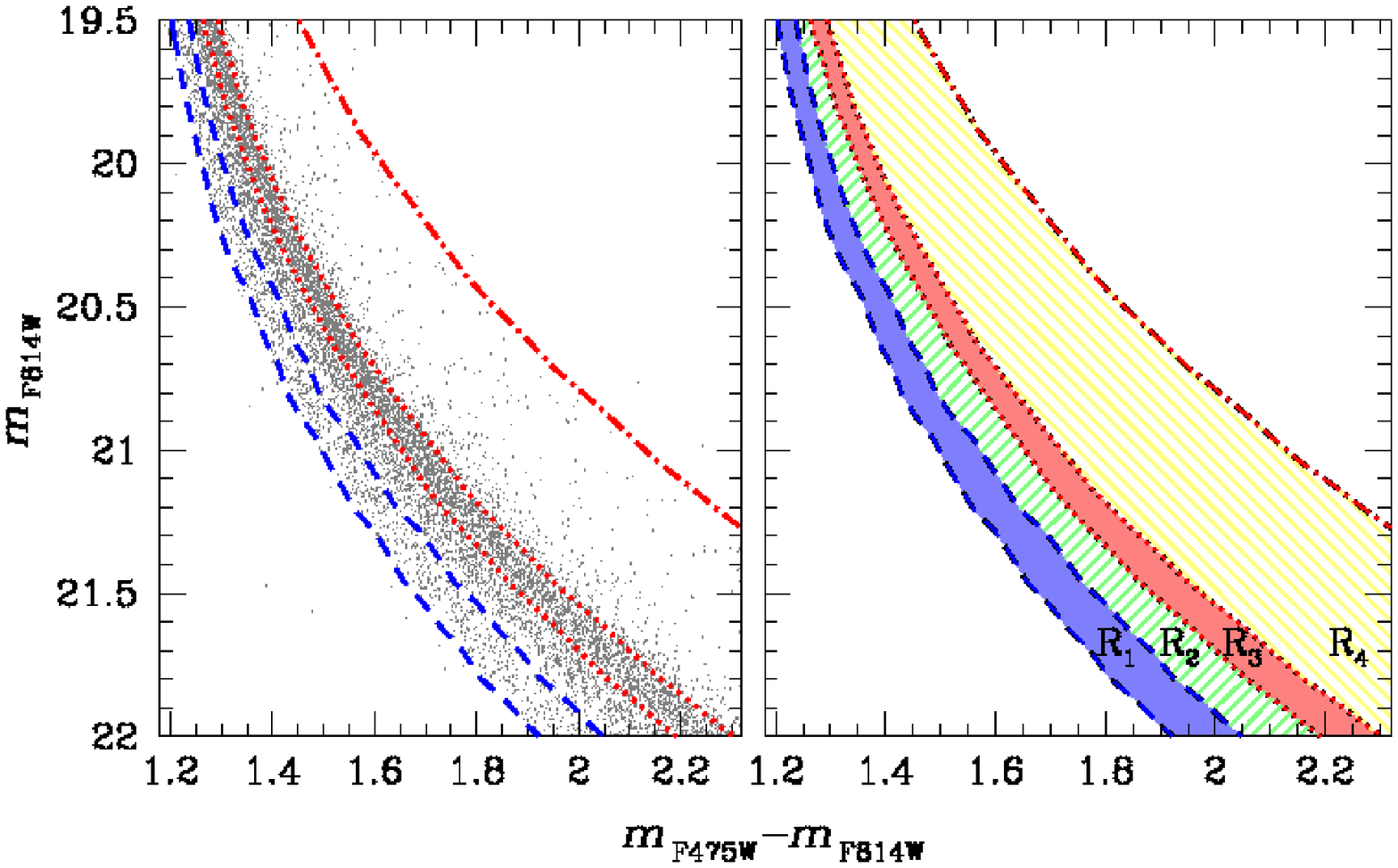}
      \caption{ \textit{Left panel:} ${\it m}_{\rm F814W}$ vs.\ ${\it
          m}_{\rm F475W}-{\it m}_{\rm F814W}$ CMD for MS stars. Blue
          dashed lines and red dotted lines are the MSRLs of the bMS
          and the rMS blue and red-shifted by $\pm$ three times
          $\sigma_{\rm bMS}$ and $\sigma_{\rm rMS}$. The red
          dashed-dotted line is displaced to the red from the fiducial
          of equal-mass rMS-rMS binaries by $+3$ times $\sigma_{\rm
          rMS}$.  These lines are the boundaries of the shaded areas
          in the \textit{right panel} that define the four regions
          used to calculate the LFs of the three MSs and the binary
          fraction (see text for details).  }

   \label{regions}
   \end{figure*}
%
   \begin{figure*}[ht!]
   \centering
   \includegraphics[width=12.5 cm]{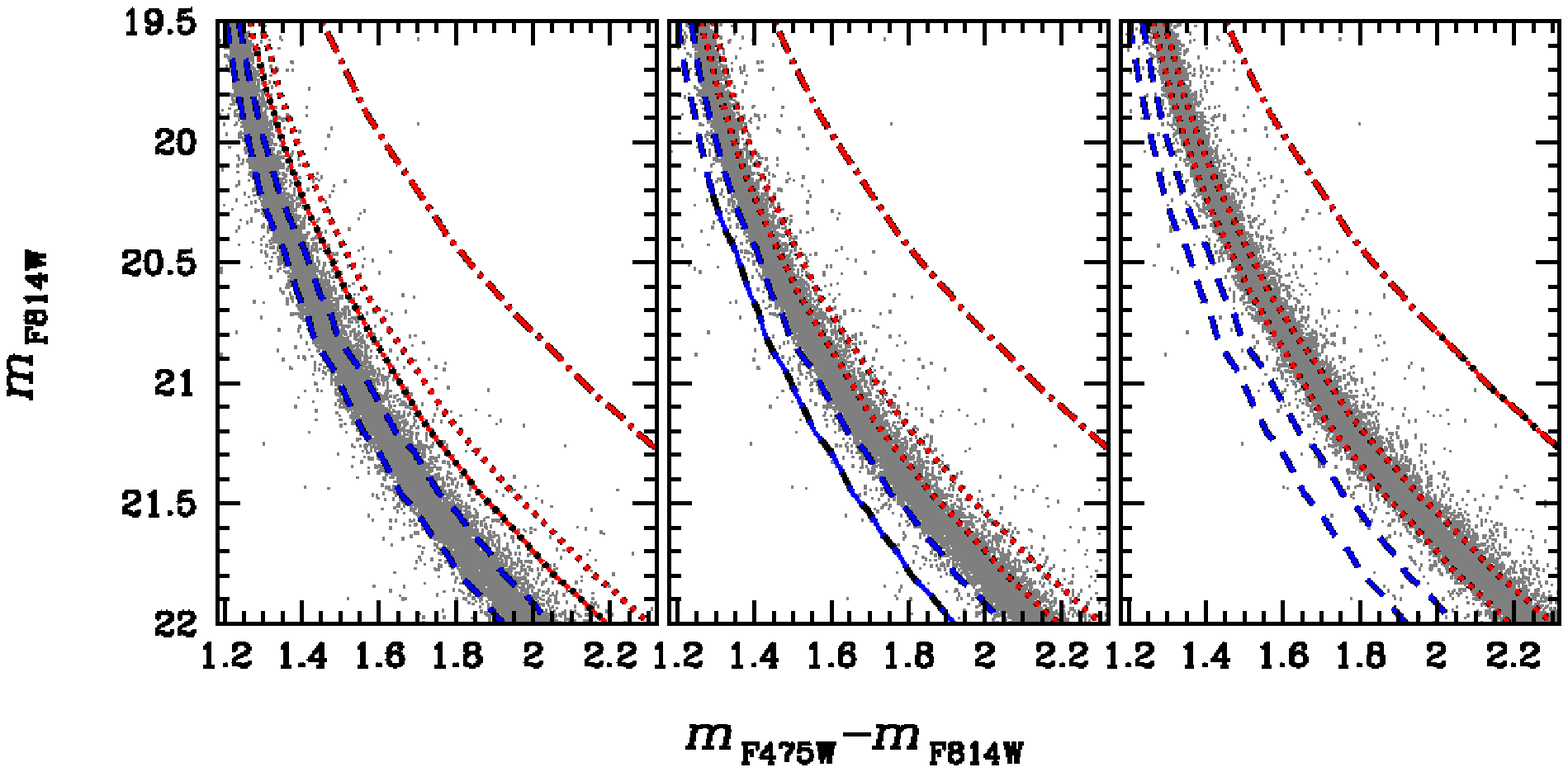}
   \caption{ Simulated CMD for single bMS (\textit{left}), mMS
            (\textit{middle}), and rMS (\textit{right}) stars.  We
            superimposed on each CMD the boundaries of the four
            regions defined in Fig.~\ref{regions}.  }
         \label{regionsAS}
   \end{figure*}
%

To illustrate our setup, the left panel of Fig.~\ref{regions} shows a
zoom of the external field CMD around the region where the division in
the MS is most evident.  The two red dotted lines are constructed by adding and
subtracting from the colour of the rMSRL a value equal to three times $\sigma_{\rm
rMS}$.  Similarly, the blue dashed lines are the bMSRL shifted to the
blue 
and red by three times $\sigma_{\rm bMS}$.  The red dot-dashed
lines mark the loci of equal-mass rMS-rMS binaries red-shifted by
$3\times \sigma_{\rm rMS}$.  The four regions delimited by these five
lines are marked in the right panel with different colour codes and
are named $R_{1}$, $R_{2}$, $R_{3}$, and $R_{4}$.

The bulk of bMS stars are located within the region $R_{1}$, while
regions $R_{2}$ and $R_{3}$ are mainly populated by mMS and rMS
stars. Region $R_{4}$ mainly contains binary systems formed by a pair
of MS stars. Obviously not all the single b(m,r)MS stars are located
within $R_{1(2,3)}$ but a fraction of them migrates to nearby
regions because of the measuring error.  In addition, all these regions are
also populated by binaries.

We use AS tests to estimate the fractions of the ``misplaced'' (and of the
correctly placed) single stars. To each region $R_{\rm i}$, we
associate
a fraction $f^{\rm rMS}_{\rm i}$ of rMS stars,
a fraction $f^{\rm mMS}_{\rm i}$ of mMS stars, and
a fraction $f^{\rm bMS}_{\rm i}$ of bMS stars.
This is illustrated in Fig.~\ref{regionsAS}, where we show the
simulated CMDs for single stars of bMS (left panel), mMS (middle
panel), and rMS (right panel).

In an analogous way we add artificial binary systems ---once assumed a binary
distribution and a mass-ratio distribution--- and infer the
fraction of binaries $f^{\rm BIN}_{\rm i}$ 
that fall within each CMD
region $R_{\rm i}$, as we later see.

Our observables are the total numbers $N_{\rm i=1,2,3,4}$ (corrected
for completeness) of stars within each region $R_{\rm i=1,2,3,4}$,
which can be expressed as the sum of four terms containing the four
unknown $N_{\rm bMS}$, $N_{\rm mMS}$, $N_{\rm rMS}$, and $f^{\rm
BIN}$, according to the four relations (for $i=1,2,3,4$)
\begin{equation}
\label{eq:1}
  N_{\rm i} = N_{\rm bMS} f^{\rm bMS}_{\rm i} + N_{\rm mMS} f^{\rm
    mMS}_{\rm i} + N_{\rm rMS} f^{\rm rMS}_{\rm i} + f^{\rm BIN}
    N_{\rm MS} f^{\rm BIN}_{\rm i},
\end{equation}
where the symbols $N_{\rm bMS}$, $N_{\rm mMS}$, and $N_{\rm rMS}$ are the
total number of single bMS, mMS, and rMS in our sample, and for
convenience we have introduced the quantity $N_{\rm MS}=N_{\rm
bMS}+N_{\rm mMS}+N_{\rm rMS}$, which is the total number of single
stars in our sample. Finally, the quantity $f^{\rm BIN}$ expresses, as
a fraction of $N_{\rm MS}$, the unknown total binary fraction.

To obtain the LFs, we need to estimate the unknowns quantities $N_{\rm
bMS}$, $N_{\rm mMS}$, $N_{\rm rMS}$, and $f^{\rm BIN}$ at different
magnitude intervals.\footnote{ Note that the LFs of the three MSs, are
indeed provided by the quantities $N_{\rm bMS}$, $N_{\rm mMS}$, and 
$N_{\rm rMS}$ obtained at different magnitudes.} As an adequate
compromise, we chose to divide the CMD into intervals of 0.25 mag in
$m_{\rm F814W}$, within which we estimated: $N_{\rm i=1...4}$,
$f^{\rm bMS}_{\rm i=1...4}$, $f^{\rm mMS}_{\rm i=1...4}$, and $f^{\rm
rMS}_{\rm i=1...4}$ for each magnitude bin.

However, there is an interplay between the $f^{\rm BIN}_{\rm i}$ and
the LFs, as the $f^{\rm BIN}_{\rm i}$ depend strongly on the
LFs and visa-versa.  To break this degeneracy, we adopted an iterative
procedure.  At the first iteration, we assumed $f^{\rm BIN}=0$ and
solve the system of four forms ($i=1,2,3,4$) of Eq.~\ref{eq:1} in the unknowns
$N_{\rm bMS}$, $N_{\rm mMS}$, and $N_{\rm rMS}$.  These numbers
provide us with the first (crude) estimates that we use to determine the
$f^{\rm BIN}_{\rm i}$.

To determine $f^{\rm BIN}_{\rm i}$, we created a simulated 
``pure binary'' CMD under the assumptions given in previous 
section (i.\ e.\ a flat {\it f(q)} and the binary components 
belonging to any of the MSs).  We draw the primary member 
of the binary from the current observed distribution of stars 
for each sequence, constructing an artificial-star catalog 
10$\times$$N_{\rm bMS}$ bMS stars, $10\times N_{\rm mMS}$ mMS 
stars, and $10\times N_{\rm rMS}$ rMS stars (the factor of 
ten providing us with higher quality statistics).  For each of these stars, 
we determine, based 
on the F475W and F814W magnitudes and the mass-luminosity 
relation of Pietrinferni et al.\ (2004), the corresponding mass
($\mathcal{M}_1$) which we then associate 
with a secondary that has a mass randomly drawn 
between 0.08 $\mathcal{M}_{\odot}$ and $\mathcal{M}_1$.  We 
then add the flux of the secondary to that of the primary and 
arrive at the magnitudes for the binary system.

\begin{figure}[ht!]
   \centering
   \includegraphics[width=8.5 cm]{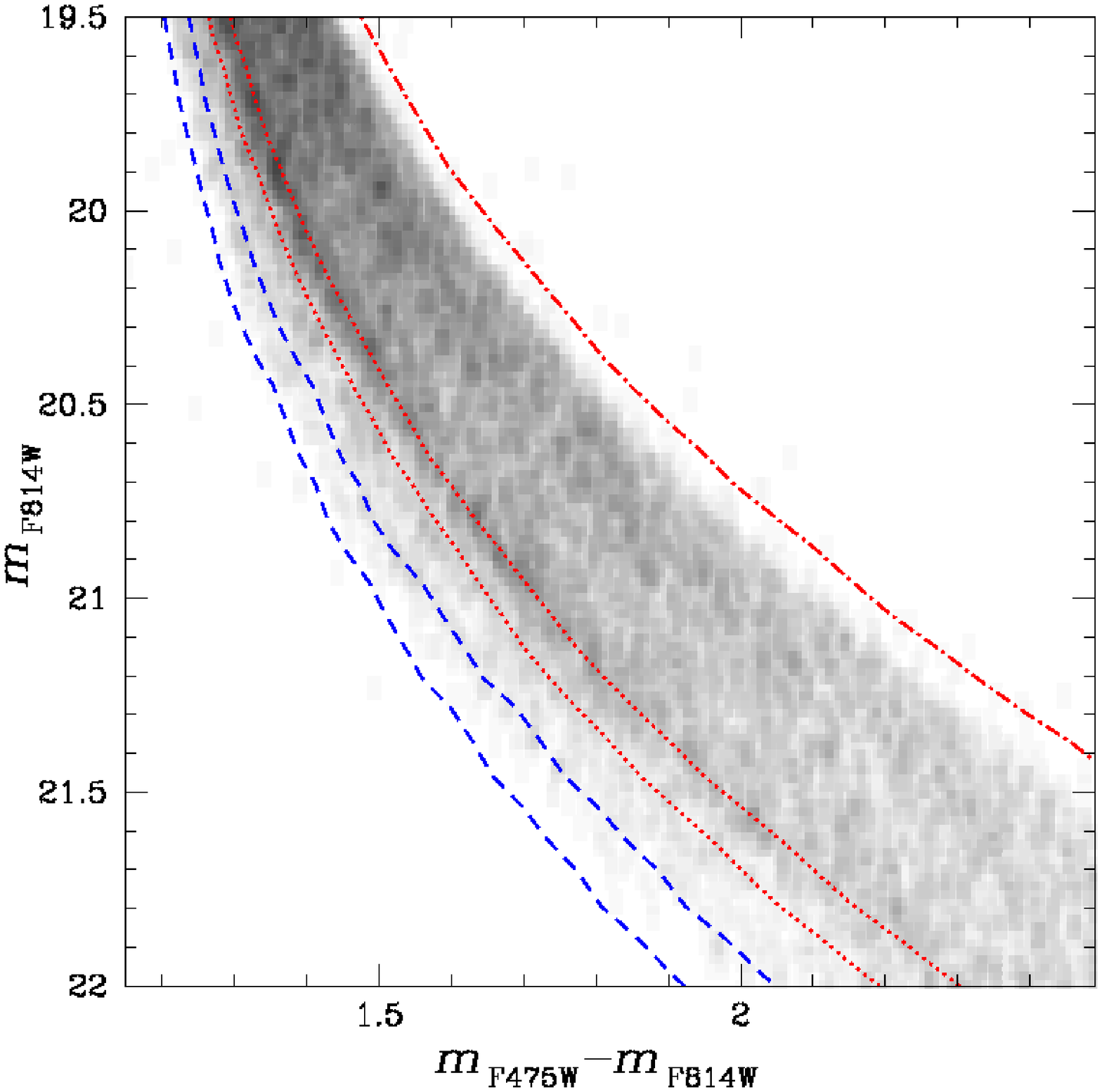}
   \caption{Hess diagram of a simulated CMD for MS-MS binaries
            alone.  We used a flat mass-ratio distribution and
            considered that the
            two components of a binary system have the same
            probability of belonging to any given MS.  We superimposed on each
            CMD the boundaries of the four regions defined in
            Fig.~\ref{regions}.  }
   \label{simu}
   \end{figure}

The Hess diagram of the resulting CMD is shown in Fig.~\ref{simu} 
with the boundaries between the four CMD regions superimposed.
This CMD can
then be used to estimate the fraction of binaries $f^{\rm BIN}_{\rm
i}$ in each region $R_{\rm i}$, by computing the ratio of the number
of binaries within each region, and the total inserted binaries.

With the values of $f^{\rm bMS}_{\rm i}$, $f^{\rm mMS}_{\rm i}$,
$f^{\rm rMS}_{\rm i}$ and a first guess of the values of $f^{\rm
BIN}_{\rm i}$ we can now solve the system of four equations
(Eq.~\ref{eq:1}) in the four unknown $N_{\rm bMS}$, $N_{\rm mMS}$,
$N_{\rm rMS}$, and $f^{\rm BIN}$.  This ends one iteration.

With the improved estimates of $N_{\rm bMS}$, $N_{\rm mMS}$, $N_{\rm
rMS}$, we simulated a new CMD, improved our estimates of $f^{\rm
BIN}_{\rm i}$, and again solved Eq.~\ref{eq:1} for $N_{\rm bMS}$, $N_{\rm
mMS}$, $N_{\rm rMS}$, and $f^{\rm BIN}$.  We iteratively repeated
the procedure until convergence is reached, i.\ e.\ until the $f^{\rm BIN}$
value changed by less than 0.001 from one iteration to the successive
one.

\begin{figure}[ht!]
   \centering
   \includegraphics[width=8.5 cm]{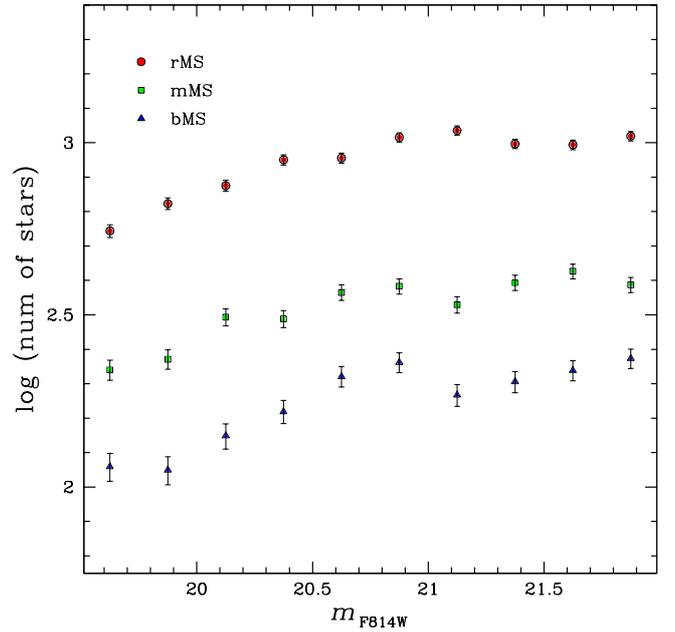}
   \caption{Luminosity function for the three MSs.}
   \label{LF}
   \end{figure}

\begin{figure*}[ht!]
   \centering
   \includegraphics[width=14.25 cm]{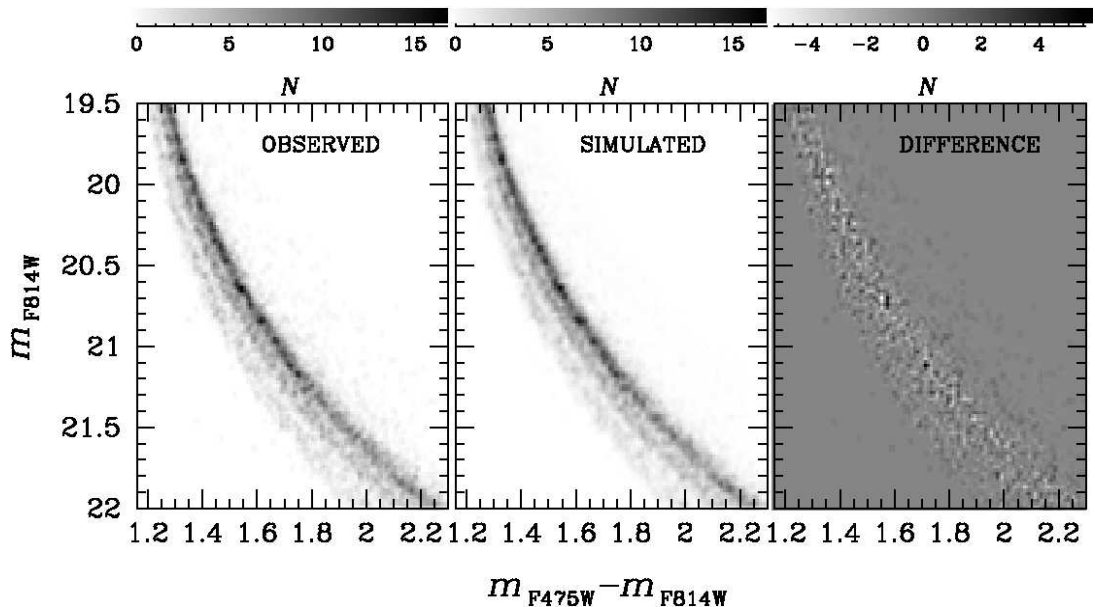}
   \caption{
            Hess diagram of the observed CMD (\textit{left}), the
            simulated CMD (\textit{middle}), and difference between
            the observed and simulated Hess diagrams (\textit{right}).
            }
   \label{simulation}
   \end{figure*}

The procedure converges to a total binary fraction for NGC 2808 of
$f^{\rm BIN} =0.045 \pm 0.011$, and to the three LFs for the three
individuals MSs displayed in Fig.~\ref{LF}.  Star counts for bMS,
mMS, and rMS for each magnitude bin are listed in 
Table~3.
The associated errors are Poisson errors and represent a lower limit
to the true uncertainties in the star counts and the binary fraction.
An inspection of this figure suggests that the three LFs have a
similar trend with magnitude; they increase up to ${\it m}_{\rm
  F814W} \sim 20.5$ and then remain nearly constant at fainter
magnitudes.  Figure~\ref{simulation} directly compares 
the observed CMD (on the left) and the simulated CMD (in the middle)
 during the last iteration. The right panel shows the difference
 between these two.

The sum over all the magnitude bins of the quantities $N_{\rm bMS}$,
$N_{\rm mMS}$, and $N_{\rm rMS}$ provides a measure of the 
population ratios among
the MSs.  The fraction of stars in one of the MSs with
respect to the total number of single MS stars ($N_{\rm MS}$) are:
($N_{\rm rMS}/N_{\rm MS}$;\ $N_{\rm mMS}/N_{\rm MS}$;\ $N_{\rm
bMS}/N_{\rm MS}$)$=$ ($62\pm2$; $24\pm2$; $14\pm3$)\%.

We note that these numbers are different from the crude estimate given in
Table~1 of Piotto et al.\ (2007), and much in closer agreement with
the fraction of (O-normal;\ O-poor;\ Super-O-poor) $=$ ($61 \pm 7$;
$22 \pm 4$; $17 \pm 4$)\% based on the RGB stars sample presented in
Carretta et al.\ (2006).

%

%

%
\begin{table*}[ht!]
\begin{center}
\caption{ Numbers of rMS, mMS, and bMS stars, corrected for
completeness, per magnitude and mass interval.  }
\begin{tabular}{ccccccc}
\hline
\hline
$\Delta {\it m}_{\rm F814W}$ & & $\Delta {\mathcal M}$& & & Number of stars & \\  
\hline
 & rMS & mMS & bMS & rMS & mMS & bMS \\ 
\hline
 19.50-19.75 & 0.751-0.770 & 0.684-0.699 & 0.611-0.623 & $ 485\pm22$ & $204\pm14$ & $103\pm10$ \\
 19.75-20.00 & 0.729-0.751 & 0.667-0.684 & 0.597-0.611 & $ 595\pm24$ & $213\pm15$ & $102\pm10$ \\
 20.00-20.25 & 0.705-0.729 & 0.646-0.667 & 0.581-0.597 & $ 671\pm26$ & $294\pm17$ & $126\pm11$ \\
 20.25-20.50 & 0.680-0.705 & 0.624-0.646 & 0.564-0.581 & $ 826\pm29$ & $294\pm17$ & $156\pm12$ \\
 20.50-20.75 & 0.654-0.680 & 0.602-0.624 & 0.544-0.564 & $ 836\pm29$ & $353\pm19$ & $196\pm14$ \\
 20.75-21.00 & 0.628-0.654 & 0.578-0.602 & 0.524-0.544 & $ 958\pm31$ & $363\pm19$ & $220\pm15$ \\
 21.00-21.25 & 0.603-0.628 & 0.556-0.578 & 0.504-0.524 & $1008\pm32$ & $323\pm18$ & $172\pm13$ \\
 21.25-21.50 & 0.579-0.603 & 0.534-0.556 & 0.484-0.504 & $ 922\pm30$ & $376\pm19$ & $192\pm14$ \\
 21.50-21.75 & 0.555-0.579 & 0.512-0.534 & 0.465-0.484 & $ 924\pm30$ & $409\pm20$ & $206\pm14$ \\
 21.75-22.00 & 0.530-0.555 & 0.489-0.512 & 0.442-0.465 & $ 967\pm31$ & $369\pm19$ & $224\pm15$ \\
\hline
\hline
\end{tabular}
\end{center}
\label{tabLFs}
\end{table*}

\section{The mass functions of the three MSs}
\label{sec:MFs}
The LFs can be transformed into MFs by using a theoretical
mass-luminosity relation.  However, we note that there is
some uncertainty in this approach, mainly related to the still
significant uncertainties that affect the color-$T_{\rm eff}$
relations and bolometric correction scales needed to transform the
model predictions from the theoretical plane to the observational one
(see the discussion in Cassisi 2007).

\begin{figure*}[ht!]
   \centering
   \includegraphics[width=8.25 cm]{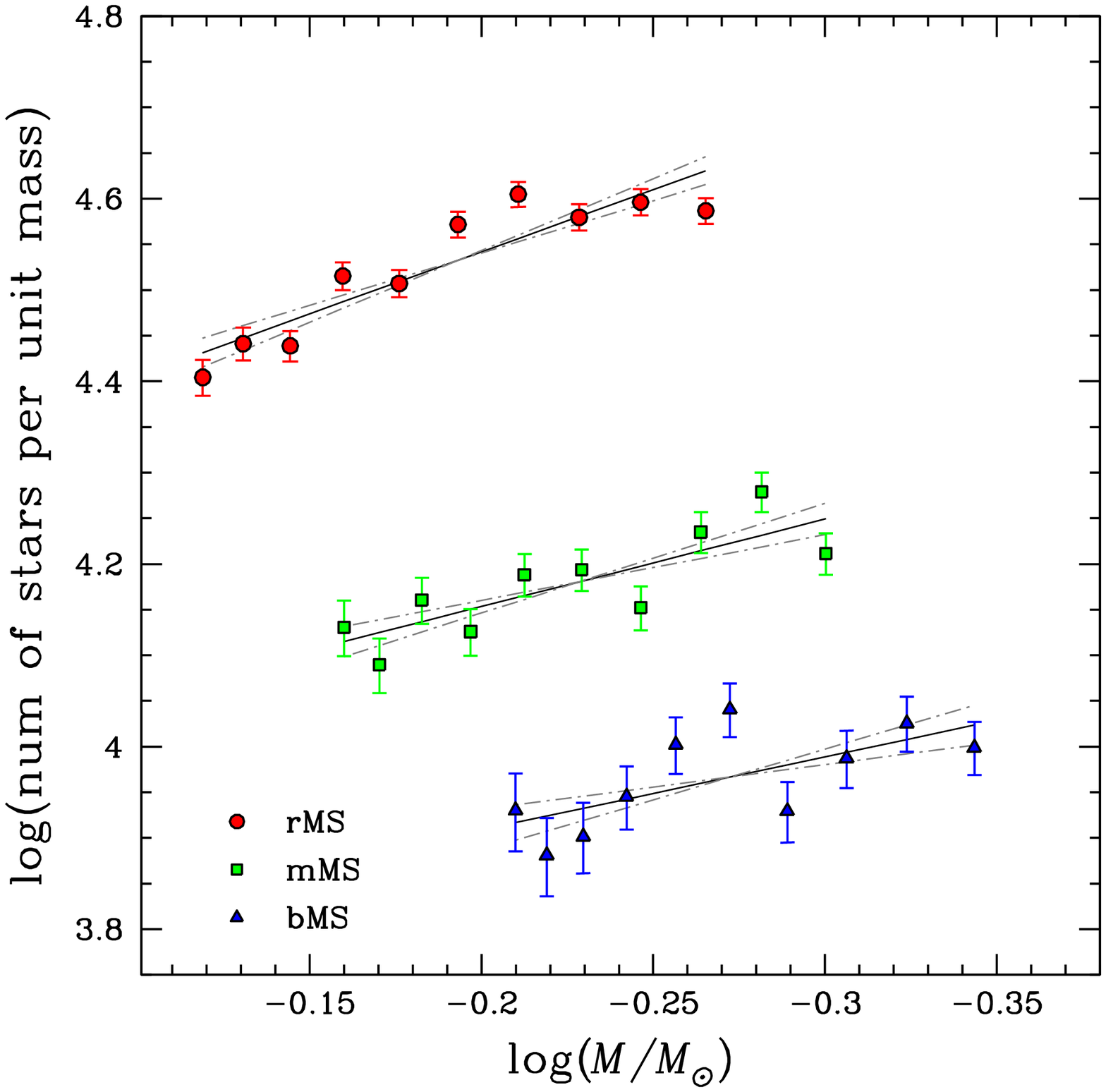}
   \includegraphics[width=8.25 cm]{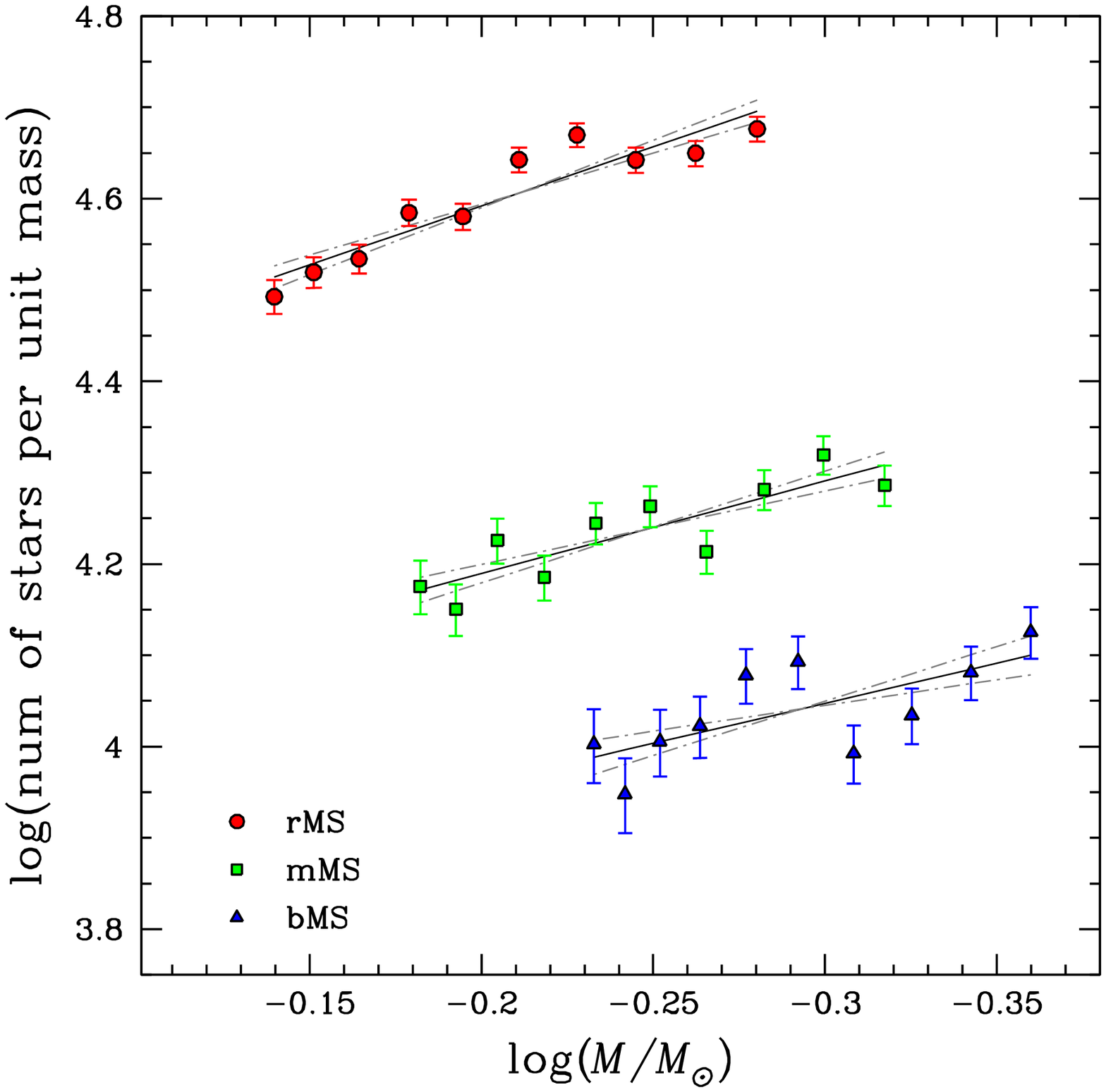}
   \caption{ Mass function for the three MSs obtained by using the
        observed LFs of Fig.~\ref{LF} and the mass-luminosity
        relations from Pietrinferni et al.\ (2004, \textit{left}) and
        Dotter et al.\ (2007, \textit{right}). Continuous black lines
        are the best-fit straight lines, while grey dotted-dash
        lines are the maximum- and minimum-slope lines.  Note that
        star counts have been done in mass intervals, by dividing the
        numbers of stars estimated in luminosity bins by the
        respective bin widths in mass. } 
         \label{MFs}
   \end{figure*}

The models we used in this paper account for the different helium
content of the three MSs and allow us to gather information on the
shape of the MFs of the three MSs of NGC 2808. In Fig.~\ref{MFs}, we
compare the MFs derived from the LFs of Fig.~\ref{LF} by using the
mass-luminosity relation of Pietrinferni et al.\ (2004) and Dotter et
al.\ (2007).  We note that the error bars represent the Poisson errors
only (derived from the Poisson error of the star counts in the LF), 
 hence
must be considered a lower limit to the true MF errors.  
 The magnitude intervals have been transformed into mass intervals
using the BaSTI mass-luminosity relations and are listed in
Table~3.  We fitted the resulting values by applying a
least squares fit of straight lines.  Best-fit lines are plotted as black
continuous lines, while gray dotted-dashed lines represent the 
minimum- and maximum-slope lines.

If we assume that the MF is described by a power law of the form ${\it
dN} / d {\mathcal M} \propto {\mathcal M}^{\rm \alpha}$, we have for
the bMS, the mMS, and the rMS the slopes ($\alpha$) of $-1.20 \pm 0.20$,
$-0.84 \pm 0.22$, and $-0.84 \pm 0.28$, respectively in the case of a
Pietrinferni et al.\ (2004) mass-luminosity relation, while the Dotter
et al.\ (2007) models give for the three MSs slightly steeper MFs with
slope of $-1.29 \pm 0.17$, $-1.01 \pm 0.21$, and $-0.88 \pm 0.32$.
These numbers indicate that there is no evidence of a significant
difference among the slopes of the three MSs, in the limited mass
interval considered in the present study.
However, we note that the MF for the rMS does appear to flatten below 
${\mathcal M} / {\mathcal M}_{\rm \odot} \sim 0.6$, at variance with the 
other two MFs.

%

\section{Discussion}

The initial MF (IMF) is a fundamental characteristic of a stellar population. 
Similarities or differences in the IMFs of the different populations in 
NGC 2808 might reflect differences in the environment
(primordial nebula with different
composition, temperature, density, etc.)  from which those populations formed.
Unfortunatly, in a GC
we can only measure the present-day MF, and often in a limited
region of the cluster.  The present-day MF can depend on several
effects, including internal dynamical evolution and dynamical
interaction with the Galactic gravitational potential (see
e.\ g.\ Piotto \& Zoccali 1999, De Marchi 2010). 

In the case of a GC with multiple
stellar generations, the situation is likely much more complicated,
because the different populations might 
evolve dynamically in quite different ways.
In particular, the models of D'Ercole et al.\
(2008) predict that a large fraction (up to 90\% or more)
first-generation stars should be lost early in the cluster evolution
because of the expansion and stripping of the cluster's outer
layers resulting from the mass loss subsequent to 
 SNe explosions. The  possibility
that first-generation stars were much more numerous at the time of 
cluster formation  could also account for the
polluting material needed to explain the He enrichment of the
second (and third) generations. 

For the first time, we heve presented the MFs of three 
different stellar generations in a GC.  Unfortunately, these MFs are 
limited to a very small mass interval ($\le0.3m_\odot$), which 
in addition to  the present-day mass function being the result 
of the cluster dynamical evolution, prevents us from drawing any conclusion 
 about the similarities and differences among the IMFs of
the different populations. We have also be unable to compare with the
present day MFs of
other clusters, because of both the limited mass interval 
sampled by our
investigation and the possibility that many (or all) other clusters host 
multiple stellar generations, which has not been identified in previous investigations. 
Here, we simply note that
the first-generation stars (which should be associated with the rMS, see
Piotto et al.\ 2007) have a MF that is quite similar to the 
 present day MFs of the other
generations, and that there is marginal evidence of a possibly higher mass at which the MF starts to deviate from
a  uniform power law (to be investigated in greater detail) compared with the
other populations. 

In any case, the MFs presented in this paper provide a first
observational constraint
on models that describe the formation and evolution of multiple
generations of stars in GCs.
 It will be important to extend these MFs to a larger mass interval.
Another missing, but important, piece of
information is the radial distribution of the different stellar
populations in NGC 2808. D'Ercole et al.\ (2008)   use hydrodynamical
model to predict quite a different distribution of
first- and second-generation stars.  Future \textit{HST\,} observations
using near-UV filters
 will be of crucial importance. The F275W and F336W bands have
proven to be extremely efficient in separating different stellar
populations in the MS (see, e.\ g.\ Bellini et al.\ 2010), whose
monitoring from the cluster center to the outskirts provides us with
information on the origin of the populations.
The recently approved Cycle 19 HST program (GO12605; PI Piotto) 
will provide us with additional data to both extend the MF mass-interval coverage, and follow the radial variation in the MFs of the 
different stellar populations, taking advantage of the appropriate 
near-UV observations.

\section{Summary}
\label{sec:conc}

We have used high precision \textit{HST} ACS/WFC
photometry to estimate the LFs of the three MSs of NGC 2808, and
measure the fraction of binaries in this GC.  The field we have analyzed is
located in the south-west quadrant, between $\sim1.5$ and $\sim6$
arcmin from the cluster center, and we have limited our study to stars with
$19.5\le {\it m}_{\rm F814W} \le 22.0$, corresponding to $\sim 0.75
\le {\mathcal M}/{\mathcal M}_{\odot} \le 0.44$.
The present investigation represents the first attempt to measure the
present day MF of different sub-populations in a Galactic GC.  The
main results of this paper can be summarised as follows:
\begin{enumerate}
\item In the studied region and luminosity (mass) interval, we
      have estimated a total fraction $f^{\rm BIN} \sim0.05$ of binaries.
\item The three MSs have very similar LFs, with the number of stars
      increasing up to ${\it m}_{\rm F814W} \sim 20.5$, and
      remaining nearly constant at fainter magnitudes.
\item We made a more accurate measurement of the fraction of stars in
      each MS-population, finding the ratio
      (rMS:mMS:bMS)=(0.62:0.24:0.14) in agreement with the ratios of
      O-abundances along the RGB.
\item We used the mass-luminosity relations from Pietrinferni et al.\
  (2004) and Dotter et al.\ (2007) to convert the observed LFs into
  MFs.
 On average, the three MFs can be 
 represented approximately by power laws.
The power law exponent, (calculated as the average of the MF
slopes in the log-log diagram obtained from the two adopted
mass-luminosity relations) is $\alpha=-1.2\pm0.3$ for the rMS,
$\alpha=-0.9\pm0.3$ for the mMS, and $\alpha=-0.9\pm0.4$ for the bMS.
On the basis of the errors, the slope differences are insignificant.
These slopes are consistent with the typical slope of the MF of
clusters with a central concentration similar to that of NGC 2808 (De
Marchi 2010).  We note that the MF of the rMS seems to deviate from a
power law, flattening below ${\mathcal M}
/ {\mathcal M}_{\rm \odot} \sim 0.6$.  This effect is present
independently of the adopted mass-luminosity relation.  Though
interesting, this result needs to be confirmed with a more extended
(to smaller masses) MF.
    \item Finally, this study provides accurate measurements of 
          the fiducial lines for the three MSs (Table~1), which 
          can be directly compared with theoretical predictions.
\end{enumerate}

\section*{APPENDIX A. Effect of the binaries mass-ratio distribution on the mass functions}
\label{appendiceA}
As already discussed in Sect.~\ref{BIN} the choice of a correct
mass-ratio distribution of stars in MS-MS binary systems is crucial
for a reliable estimate of the binary fraction.  
We have assumed in this paper that the binaries in NGC 2808 have 
flat distributions in {\it f}({\it q}), consistent with the 
findings of Milone et al.\ (2011) in their study of MS-MS binaries 
in 59 GCs.
In this Appendix, we investigate whether a different mass-ratio
distribution could effect the main results of this paper.

\begin{figure}[ht!]
   \centering
   \includegraphics[width=8.5 cm]{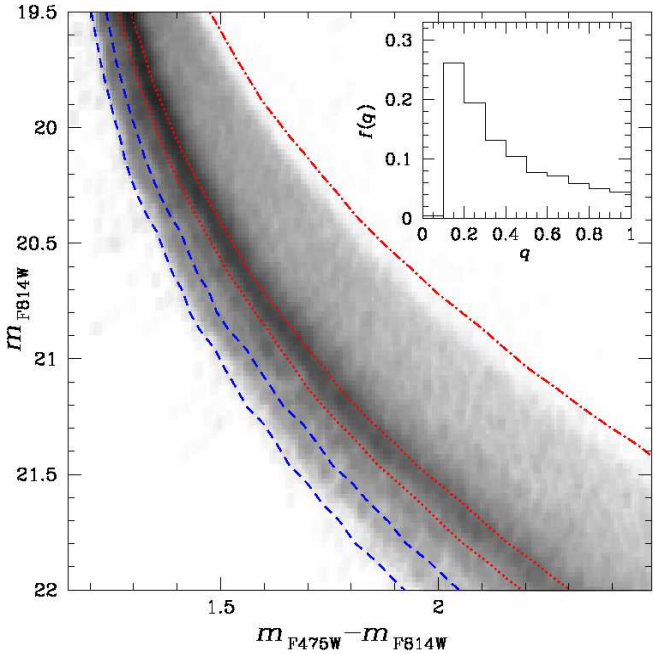}
   \caption{ Hess diagram of a simulated CMD made of MS-MS binaries
     \textit{only}. The adopted mass-ratio distribution is shown in
     the inset.  }
   \label{simurandom}
   \end{figure}

Tout (1991) suggested that {\it f}({\it q}) can be derived by randomly
extracting secondary stars from the observed initial MF.  The Hess
diagram of a CMD made of binary systems only obtained by randomly
extracting pair of stars from the MFs of NGC 2808 is plotted in
Fig.~\ref{simurandom} and the corresponding mass ratios are displayed
in the inset.

It should be noted that Sollima et al.\ (2007), on the basis of their
study of the binaries population in 13 GCs, suggested that the
mass-ratio distribution derived from the above procedure could
significantly increase the fraction of binaries with low-mass
secondaries and produce a significant overestimate of the binary
fraction.
Furthermore, as said, the Tout (1991) distribution is also in 
sharp disagreement with the observed distributions of binaries 
with $q > 0.5$ presented by Milone et al.\ (2011) for a sample 
of 59 GCs.
While a precise measurement of {\it f}({\it q}) for NGC 2808 
binaries is beyond the purpose of this paper 
(and is not even possible with the given data set),
we emphasise that in the
case of binary stars formed by random associations between stars of
different masses we expect a higher fraction of  binaries with
  {\it  q}$<$0.5 
than can be found with the present data set.
%
%

\begin{figure}[ht!]
   \centering
   \includegraphics[width=8.5 cm]{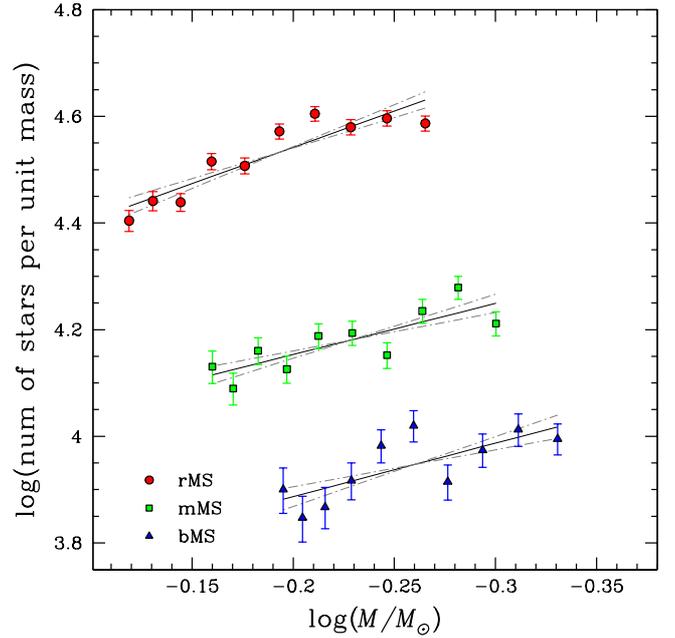}
   \caption{ Mass function of the three MSs obtained by using the
     BaSTI mass-luminosity relations and assuming the mass-ratio
     distribution of MS-MS binaries shown in Fig.~\ref{simurandom}.  }
   \label{mfrandom}
   \end{figure}

Once we had fixed the {\it f}({\it q}) shape, we followed the
procedure of Sects.~\ref{LUFU} and~\ref{sec:MFs} to estimate the
fraction of binaries and the MFs of the three MSs.  For simplicity, we
used BaSTI models only.  As expected, we found an higher $f^{\rm BIN} =
0.09 \pm 0.01$ and obtained the MFs plotted in Fig.~\ref{mfrandom}.
The three MFs are slightly steeper, but fully consistent within  the
uncertainties, with those presented in Sect.~\ref{sec:MFs}.  In this
case, we find  that the rMS, the mMS, and the bMS MF have slopes of
$-1.36\pm0.24$, $-0.96\pm0.28$, and $-1.01\pm0.34$, respectively.
These results, obtained for an `extreme' {\it f}({\it q}) demonstrate
that the adopted binaries mass-ratio distribution has a small effect
on the obtained MFs and does not change the main conclusions of this
paper.

\section*{APPENDIX B. Effect of the binaries on  the mass functions}
\label{appendiceB}
\begin{figure}[!hp]
   \centering
   \includegraphics[width=8.5 cm]{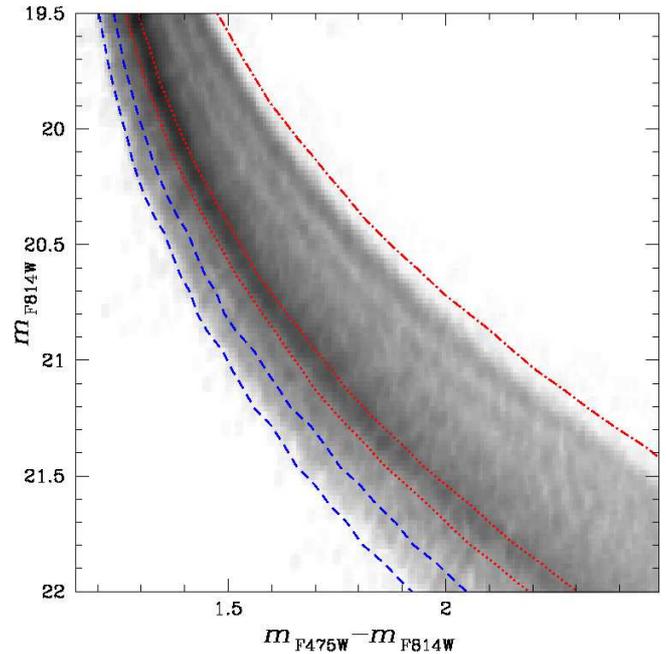}
   \caption{Hess diagram of a simulated CMD made of MS-MS binaries
               \textit{only}.  
               In contrast to the similar plot of Fig.~\ref{simu}, 
               in this case we assumed that both
               components of a binary systems come from the same MS. }
   \label{simurandom0}
   \end{figure}

In this work, we have assumed that both components of the binary
systems have the same probability of belonging to any MS.  As anticipated in
the text, we explored a different extreme scenario to
investigate how the membership of binaries to different MSs affects the derived MFs and the binary
fraction. In this scenario, we assume that both the components, in
every binary system, are taken from the same stellar population, i.\ e.\
from the same MS.  The corresponding Hess diagram for a CMD made of
binary systems only, is shown in Fig.~\ref{simurandom0}.

We adopted the same procedure described in Sects.~\ref{LUFU}
and~\ref{sec:MFs} to estimate the fraction of binaries and the MFs of
the three MSs. For simplicity, we used in this case only BaSTI
models.  The resulting value of $f^{\rm BIN}=0.05\pm0.01$ is very
close to the one obtained in Sect.~\ref{sub:LF} and the rMS, the mMS, 
and the bMS MF have similar slopes of $-1.25\pm0.22$, $-0.87\pm0.25$, 
and $-0.91\pm0.32$, respectively, again similar to those of 
Sect.~\ref{sec:MFs}.  Once again, these results suggest that the 
procedures adopted in this paper are insignificantly affected by 
the adopted scenario.

\begin{acknowledgements}
APM, GP, SC, and AFM acknowledge support of ASI and INAF under the
programs ASI-INAF n. I/009/10/0 and PRIN-INAF 2009.  
JA acknowledges the support of STScI grant GO-10922.
\end{acknowledgements}
%
%
\bibliographystyle{aa}

\end{document}